\documentclass[reprint,nofootinbib,amsmath,amssymb,showkeys,aps]{revtex4-1}

\usepackage{amssymb}
\usepackage{array,multirow}
\usepackage{enumerate}
\usepackage{bm}
\usepackage{amsmath}
\usepackage{graphicx}
\usepackage{xcolor}

\usepackage[flushleft]{threeparttable}

\begin{document}

\title{Late-time phenomenology required to solve the $H_0$ tension in view of the cosmic ladders and the anisotropic and angular BAO data sets}

\author{Adri\`a G\'omez-Valent$^{1,2}$}\email{agvalent@roma2.infn.it}
\author{Arianna Favale$^{1,2}$}\email{afavale@roma2.infn.it}
\author{Marina Migliaccio$^{1,2}$}\email{migliaccio@roma2.infn.it}
\author{Anjan A. Sen$^{3}$}\email{aasen@jmi.ac.in}

\affiliation{$^1$ Dipartimento di Fisica, Università di Roma Tor Vergata, via della Ricerca Scientifica, 1, 00133, Roma, Italy}
\affiliation{$^2$ INFN, Sezione di Roma 2, Università di Roma Tor Vergata, via della Ricerca Scientifica, 1, 00133 Roma, Italy}
\affiliation{$^3$ Centre for Theoretical Physics, Jamia Millia Islamia, New Delhi-110025, India}

\begin{abstract}
The $\sim 5\sigma$ mismatch between the value of the Hubble parameter measured by SH0ES and the one inferred from the inverse distance ladder (IDL) constitutes the biggest tension afflicting the standard model of cosmology, which could be pointing to the need of physics beyond $\Lambda$CDM. In this paper we study the background history required to solve the $H_0$ tension if we consider standard prerecombination physics, paying special attention to the role played by the data on baryon acoustic oscillations (BAO) employed to build the IDL. We show that the anisotropic BAO data favor an ultra-late-time (phantom-like) enhancement of $H(z)$ at $z\lesssim 0.2$, accompanied by a transition in the absolute magnitude of supernovae of Type Ia $M(z)$ in the same redshift range. This agrees with previous findings in the literature. The effective dark energy (DE) density must be smaller than in the standard model at higher redshifts. Instead, when angular BAO data (claimed to be less subject to model dependencies) is employed in the analysis, we find that the increase of $H(z)$ starts at much higher redshifts, typically in the range $z\sim 0.5-0.8$. In this case, $M(z)$ could experience also a transition (although much smoother) and the effective DE density becomes negative at $z\gtrsim 2$. Both scenarios require a violation of the weak energy condition (WEC), but leave an imprint on completely different redshift ranges and might also have a different impact on the perturbed observables. They allow for the effective crossing of the phantom divide. Finally, we employ two alternative methods to show that current data from cosmic chronometers do not exclude the violation of the WEC, but do not add any strong evidence in its favor neither. Our work puts the accent on the utmost importance of the choice of the BAO data set in the study of the possible solutions to the $H_0$ tension.  
\end{abstract}


\maketitle


\section{Introduction}\label{sec:intro}

It is not possible to constrain the Hubble parameter, $H_0$, with uncalibrated data on supernovae of Type Ia (SNIa) or baryon acoustic oscillations (BAO). One needs to calibrate first these data sets with a measurement of the absolute magnitude of SNIa, $M$, and the comoving sound horizon at the baryon-drag epoch, $r_d$, respectively. These are the calibrators of the so-called direct and inverse cosmic distance ladders, which are two of the main ways we have of measuring the current expansion rate of the Universe. The latter, despite not fundamental, is of course a very important quantity, since it enters the computation of cosmic times and distances. 

The SH0ES Team has measured $M^{R22} = (-19.253 \pm 0.027)$ mag using calibrated Cepheid variable stars in galaxies that also host SNIa. This leads to the measurement of $H_0^{R22}=(73.04\pm 1.04)$ km/s/Mpc making use of the supernovae in the Hubble flow \cite{Riess:2021jrx}. On the other hand, the {\it Planck} Collaboration finds $r^{P18}_d=(147.09\pm 0.26)$ Mpc and $H^{P18}_0=(67.36\pm  0.54)$ km/s/Mpc under the assumption of $\Lambda$CDM and using the TT,TE,EE+lowE+lensing cosmic microwave background (CMB) likelihood \cite{Planck:2018vyg}. This implies a $\sim 5\sigma$ mismatch between the local measurement by SH0ES and the {\it Planck}/$\Lambda$CDM inference. A similar level of tension with SH0ES is found when $r^{P18}_d$ is employed to calibrate the anisotropic (or 3D) BAO data in fitting analyses of the standard model (see e.g. \cite{DES:2017txv}) or by using these calibrated data with SNIa apparent magnitudes in parametric and cosmographical analyses \cite{Aubourg:2014yra,Cuesta:2014asa,Lemos:2018smw,Feeney:2018mkj,DES:2018rjw}. We refer the reader to the reviews \cite{Verde:2019ivm,Perivolaropoulos:2021jda,Riess:2023egm} for further details.

\begin{figure}
    \centering
    \includegraphics[scale=0.38]{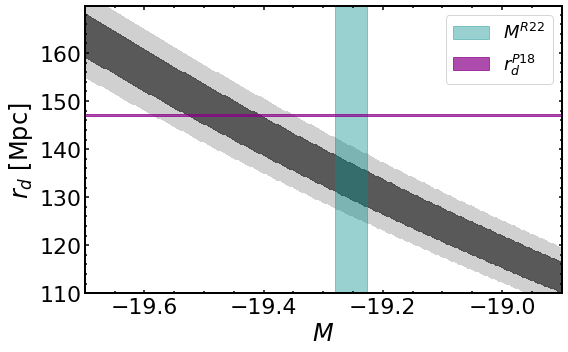}
    \caption{Degeneracy band at $68\%$ and 95$\%$ C.L. in the $M$-$r_d$ plane (in grey) inferred from uncalibrated BAO and SNIa data. We have obtained these constraints using the method previously employed in \cite{Gomez-Valent:2021hda}, which is based on the Index of Inconsistency by Lin and Ishak \cite{Lin:2017ikq}, see Appendix \ref{sec:appendixA} for details. We also include the vertical band with the SH0ES measurement of $M$ ($M^{R22}$) \cite{Riess:2021jrx} and the horizontal band with the $\Lambda$CDM value of $r_d$ obtained by {\it Planck} ($r_d^{P18}$) \cite{Planck:2018vyg}, both at 1$\sigma$ C.L. It is evident that the intersection of these two bands lie far away from the region preferred by the uncalibrated 3D BAO and SNIa data.}
    \label{fig:degMrd}
\end{figure}

The discrepancy between the two cosmic ladders in the context of the standard model could be due to some issue with their calibrators \cite{Bernal:2016gxb,Aylor:2018drw,Camarena:2019moy,Gomez-Valent:2021hda}, either by unaccounted for systematic errors or by new physics. In order to grasp the possible origin of the Hubble tension it is useful to analyze the constraints in the $M-r_d$ plane that are obtained by studying the compatibility of uncalibrated SNIa and 3D BAO data. This forces the calibrators of the direct and inverse distance ladders to lie in a quite narrow degeneracy band, see the grey contours in Fig. \ref{fig:degMrd}. In that figure we also plot the SH0ES constraint $M^{R22}$ (in cyan) and the {\it Planck}/$\Lambda$CDM constraint $r_d^{P18}$ (in purple). The mismatch between the overlapping region of these two bands and the SNIa+BAO degeneracy band is nothing else but the clear manifestation of the Hubble tension.

In the absence of systematics in the data, one possibility to try to solve the tension is to consider a departure from the standard model before the decoupling of the CMB photons, capable of decreasing the distance traveled by the sound waves in the photo-baryon fluid\footnote{This has to be complemented also by changes in the late Universe with respect to the {\it Planck}/$\Lambda$CDM best-fit cosmology, of course. To understand why see e.g. \cite{Jedamzik:2020zmd}.}. This can be achieved in different ways: by changing the strength of gravity \cite{SolaPeracaula:2019zsl,Ballesteros:2020sik,SolaPeracaula:2020vpg,Braglia:2020iik,Braglia:2020auw,Benevento:2022cql,SolaPeracaula:2023swx,Gomez-Valent:2023hov,Kable:2023bsg}, modifying the recombination time with the help of primordial magnetic fields \cite{Jedamzik:2020krr} or varying atomic constants \cite{Liu:2019awo,Hart:2019dxi,Sekiguchi:2020teg,Lee:2022gzh}, altering the shape of the primordial power spectrum \cite{Hazra:2022rdl}, or by considering different forms of early dark energy \cite{Karwal:2016vyq,Poulin:2018cxd,Agrawal:2019lmo,Lin:2019qug,Gomez-Valent:2021cbe,Niedermann:2019olb,Niedermann:2021vgd,Gomez-Valent:2022bku,Cruz:2023lmn} or dark radiation \cite{Bernal:2016gxb}. See the reviews \cite{Schoneberg:2021qvd,DiValentino:2021izs,Abdalla:2022yfr} for a more extended bibliography. These options are appealing, but it is important to mention that their ability to loosen the tension is in general limited by e.g. their impact on photon diffusion \cite{Knox:2019rjx} or the early integrated Sachs-Wolfe effect \cite{Vagnozzi:2021gjh}, and typically require much larger values of the spectral index of the primordial power spectrum of scalar perturbations \cite{Poulin:2018cxd,SolaPeracaula:2020vpg,Vagnozzi:2021gjh,Kable:2023bsg}, closer to $n_s\sim 1$. This is welcome by small-scale CMB experiments as the Atacama Cosmology Telecope (ACT), but not by {\it Planck} \cite{Giare:2022rvg,Calderon:2023obf,Giare:2023xoc}, especially when CMB polarization data is included in the fitting analyses, see e.g. \cite{Bernal:2016gxb,SolaPeracaula:2023swx,Gomez-Valent:2023hov}.

Fig. \ref{fig:degMrd} tells us that if there is no new physics before recombination, i.e. if $r_d\sim 147$ Mpc, the absolute magnitude of SNIa, considered to be constant in time, must be $M\sim -19.4$ mag. This value is in clear tension with the one measured by SH0ES in the second rung of the direct distance ladder (at $z\lesssim 0.02$). According to this very well-known result and taking for granted the absence of systematics in the SH0ES measurement of $M$, {\it Planck} and the SNIa data sets, it seems that there are only three possible routes to try to explain the $H_0$ (or $M$) tension with low-$z$ solutions:

\begin{enumerate}
\item {\it Systematics in the anisotropic BAO data:} The 3D BAO data could be affected by biases due e.g. to some model dependence introduced in the reconstruction of the BAO peak. It is therefore interesting to explore alternative BAO data sets, as the transversal or angular BAO (a.k.a 2D BAO) data set, which is perhaps less subject to model dependencies \cite{Sanchez:2010zg}, see also \cite{deCarvalho:2021azj,Nunes:2020hzy}. This could open the window to solutions that keep a large value of $M\sim M^{R22}$ in a wider redshift range and introduce some new physics at $z\gtrsim2$ to keep the location of the first peak of the CMB temperature power spectrum unaltered. This would move the grey band upwards in Fig. \ref{fig:degMrd}, towards the overlapping region of SH0ES and {\it Planck}. 

In this work we will study how the low-redshift solution to the Hubble tension changes when the 2D BAO data set is considered instead of the 3D BAO data set. The work \cite{Camarena:2019rmj} already reported the existing tension between these two data sets and showed that there is a better agreement between SH0ES and the 2D BAO data, considering a constant $M$ (see also \cite{Bernui:2023byc}). Here we will show what is the shape of $H(z)$ required not to spoil the description of the CMB, and the possible implications for the effective dark energy fluid. In addition, we will see that the data does not exclude a mild evolution of $M(z)$. 

\item {\it A sudden transition in $M$ at some point of the second rung of the direct distance ladder (at $z\lesssim 0.01$) leading to $M\sim -19.4$ mag in the Hubble flow:} This would automatically reestablish the concordance between the SH0ES measurement and the value $H_0\sim 67.5$ km/s/Mpc obtained from the fit of the $\Lambda$CDM to the 3D BAO and {\it Planck} 2018 data, essentially erasing the Hubble tension. The study \cite{Perivolaropoulos:2022khd} does not exclude this possibility, but does not find any compelling evidence in its favor neither, as it is clear from Fig. 16 of that reference. There is actually room for this transition to happen at higher redshifts and this is a possibility that we will discuss below, in point 3. 

Such a rapid transition could be due, for instance, to some difference of color correction between the SNIa in the calibration and cosmological samples, or to another unaccounted for refinement of the SNIa standardization procedure (the so-called Tripp calibration \cite{Tripp:1997wt}) \cite{Rigault:2014kaa,Perivolaropoulos:2021bds,Wojtak:2022bct,Perivolaropoulos:2022khd,Wojtak:2023sts}. The ultimate physical origin of these effects remains unknown, though. 

Alternatively, this transition in $M$ could also be caused by an ultra-late-time change of $G$ achieved through some modified theory of gravity \cite{Marra:2021fvf,Alestas:2021nmi,Perivolaropoulos:2022vql,Ruchika:2023ugh}, see also \cite{Amendola:1999vu,Wright:2017rsu}. Nevertheless, it seems unnatural that such a transition in redshift, at both cosmological and local scales, leads to Newton's constant in the local environment, exactly as in the cosmological regime at higher redshifts. We would expect this transition to alter the gravitational strength entering the Friedmann equation, therefore changing the $\Lambda$CDM background evolution and the ability of the model to fit the CMB and the anisotropic BAO data\footnote{Notice that $H_0^2=8\pi G\rho_c^0/3$. Hence, if we keep $H_0$ as in the {\it Planck}$/\Lambda$CDM model and consider a $G$ different from Newton's constant we need to change also the critical energy density $\rho_c^0$ to properly fit the 3D BAO and CMB data. The only way of doing so by keeping $\Omega_m=\rho_m^0/\rho_c^0$ to the value preferred by the SNIa data is by changing, at least, the matter density $\rho_m^0$. This would alter the amplitude and location of the CMB peaks.}. Explaining the phenomenology as due only to local (screening) effects is also contrived, since one would expect them to be present also at higher redshifts, leading to a more stable value of $M$. A natural implementation of these ideas might not be easy\footnote{However, in this work we will not focus on the ultimate physical cause of the modifications of $H(z)$ required to solve the Hubble tension. Our study will be mainly phenomenological, leaving a more theoretical analysis for the future.}.  

\item {\it A smoother transition in $M$ together with an increase of $H(z)$ happening at $z\lesssim 0.2$:} If we assume, as in the previous scenario, that the 3D BAO data are not affected by significant biases, but instead assume that the value of $M$ at the beginning of the third rung of the cosmic distance ladder is given by the SH0ES measurement, how should we change the shape of $H(z)$ to explain the CMB and 3D BAO observations while keeping $H^{R22}_0\sim 73$ km/s/Mpc? Two conditions have to be satisfied: (i) $H(z)$ has to increase at small redshifts to reach the region of $H_0$ preferred by SH0ES; and (ii) at some redshift, $H(z)$ has to go below the $\Lambda$CDM curve preferred by {\it Planck} in order to compensate the previous increase and leave the value of the angular diameter distance to the last scattering surface $D_A(z_*)$ intact. These modifications must respect the good description of the 3D BAO data. In addition, taking into account that the latter, when calibrated with $r_d^{P18}$, lead to angular diameter distances and values of $H(z)$ in good agreement with the standard model, we expect our result for $H(z)$ to automatically force the redshift evolution of $M$ in order to match the low-$z$ and high-$z$ estimates of this quantity. Otherwise, the inverse distance ladder keeps the tension between CMB and SH0ES high at the level of $M$, even if the model is able to produce a value close to $H_0^{R22}$. Therefore, we need to consider, on top of a modified Hubble expansion, a variation in $M$. This is a very important result, firstly noted in \cite{Alestas:2020zol} and further explored in \cite{Alestas:2021luu}. If the inverse distance ladder (built with 3D BAO) and SH0ES are free from systematic errors, in the context of standard pre-recombination physics, we cannot avoid the variation of $M$, regardless of the nature of the late-time new physics required at cosmological level to solve the Hubble tension.

The data still give room for this variation to happen at $z>0.01$. We will see in Sec. \ref{sec:results} that the transition in $H(z)$ and $M(z)$ is allowed to happen at $z\lesssim 0.2$ at most. Apart from that, we will also assess in this work the real need of a crossing of the phantom divide of the effective dark energy fluid. We will show that there is no clear evidence for a deviation from a cosmological constant before the transition. Our conclusions are in some sense aligned with \cite{Alestas:2020zol,Heisenberg:2022gqk,Heisenberg:2022lob}. The aim is to proceed in a model-independent fashion to try to constrain the shape of $H(z)$ at low redshifts using 3D BAO data calibrated with the measurement of $r_d$ from {\it Planck} 2018, the SH0ES prior on $H_0$ and the CMB distance prior on $D_A(z_*)$ from {\it Planck}. We will also study the impact of cosmic chronometers (CCH). The reconstruction of $H(z)$ will be used then to reconstruct the shape of $M(z)$ needed to avoid the inverse distance ladder bottleneck.

A few comments about other references on these matters are now in order. Based on \cite{Mortonson:2009qq}, in \cite{Benevento:2020fev,Camarena:2021jlr,Efstathiou:2021ocp} the authors explored transitions happening at $z=0.1$ in the context of models with a change from a cosmological constant to phantom dark energy. They found that although these models can alleviate the Hubble tension, they are unable to reconcile the values of $M$ measured with the direct and inverse distance ladders. This presumably happens too in the scenarios analyzed in e.g. \cite{DiValentino:2017rcr,Teng:2021cvy}. Similar transitions, from quintessence to phantom, were also studied in \cite{DiValentino:2020naf,Adil:2023exv}, and in \cite{Farhang:2023hen} the authors explored a modified gravity parametrization with a transition at $z=0.1$ as well and found evidence in its favor, with an increase of $H_0$. However, they used only BAO and CMB data to constrain their model. The inclusion of SNIa data in their analysis would produce exactly the same problems found in \cite{Benevento:2020fev,Camarena:2021jlr,Efstathiou:2021ocp}, since the resulting value of $M$ would be at odds with $M^{R22}$. These works do not solve the $M$ tension, since they did not consider the redshift evolution of this quantity. 

\end{enumerate}

As already mentioned in the first point, in this work we will study how the shape of $H(z)$ required to solve the Hubble tension changes with the BAO data set, assuming standard physics before recombination. In particular, we will show that the shape preferred by the 2D BAO data set is quite different from the one preferred by the 3D BAO data. In the former case, the deviations from the $\Lambda$CDM appear also at much higher redshifts ($z\gtrsim 2$) and the transition is much smoother. In addition, we will also reconstruct the shape of the deceleration parameter.

This work is organized as follows. In Sec. \ref{sec:method} we explain the methodology used in the reconstruction of $H(z)$ and $M(z)$, as well as the data sets employed in our analyses. We also describe two methods to test the violation of the weak energy condition (WEC) with the help of cosmic chronometers. In Sec. \ref{sec:results} we present the results obtained using the anisotropic and angular BAO data sets in combination with CMB priors, with and without the addition of cosmic chronometers. We also discuss in detail the violation of the WEC required by these low-$z$ solutions. Finally, in Sec. \ref{sec:conclusions} we provide our conclusions. Appendices \ref{sec:appendixA}-\ref{sec:appendixD} complement the content of the main body of the paper.


\section{Methodology and data}\label{sec:method}

\subsection{Fitting function for $H(z)$}\label{sec:Hmethod}

We assume throughout this paper a flat Friedmann-Lema\^itre-Robertson-Walker (FLRW) Universe and use the following fitting expression for the Hubble function,

\begin{equation}\label{eq:H}
H(z) = \left\{
        \begin{array}{ll}
             \bar{H}(z)+ \delta H_1(z)& {\rm if}\quad 0<z \leq z_{\rm p} \\

             \bar{H}(z)+ \delta H_2(z)& {\rm if} \quad z_p<z < z_{\rm max} \\
            H_\Lambda(z)& {\rm if} \quad z \geq z_{\rm max}
        \end{array}
    \right.
\end{equation}
with

\begin{equation}\label{eq:Hbar}
\bar{H}(z)\equiv \bar{H}_0\sqrt{1+\bar{\Omega}_m[(1+z)^3-1]}\,,
\end{equation}

\begin{equation}\label{eq:HLambda}
 H_\Lambda(z)\equiv \tilde{H}_0\sqrt{1+\tilde{\Omega}_m[(1+z)^3-1]+\tilde{\Omega}_r[(1+z)^4-1]}\,,
\end{equation}
and

\begin{equation}\label{eq:deltaH}
\delta H_1(z)\equiv a+bz+cz^2\quad ; \quad \delta H_2(z)\equiv d+ez+fz^2\qquad
\end{equation}
It is a versatile fitting expression, which can reproduce phantom and quintessence behaviors, and also permits the crossing of the phantom divide\footnote{For more complicated forms of $H(z)$ see e.g. \cite{Akarsu:2022lhx}. Our fitting function allows us to study the problem at hand with a minimal set of parameters.}. It allows us to avoid the use of any DE or modified gravity model, and also the choice of how to split the energy budget of the dark sector at late times, while sticking to $\Lambda$CDM at higher redshifts. 

The parameters $(\bar{\Omega}_m,\bar{H}_0)=(0.3153,67.36)$ entering Eq. \eqref{eq:Hbar} are fixed to the best-fit values obtained in the TT,TE,EE+lowE+lensing $\Lambda$CDM analysis by {\it Planck} \cite{Planck:2018vyg}. The quadratic polynomials $\delta H_1(z)$ and $\delta H_2(z)$ in the first and second rows of Eq. \eqref{eq:H} (given in Eq. \eqref{eq:deltaH})  parametrize deviations with respect to the mean Hubble function in the $\Lambda$CDM at $z<z_{\rm max}$\footnote{In Appendix \ref{sec:appendixB} we explore a similar parametrization of $\delta H_i(z)$, a second order polynomial in terms of $a-1$ instead of $z$, with $a=(1+z)^{-1}$ the scale factor. More tests on the robustness of our fitting function are performed in Appendix C.}. All the freedom of the model in this redshift range is transferred to the parameters $\{a,b,c,d,e,f\}$. For the part of $H(z)$ at $z\geq z_{\rm max}$ (see Eq. \eqref{eq:HLambda}) we use a Gaussian prior from {\it Planck} on $(\tilde{\Omega}_m,\tilde{H}_0)$, taking into account their correlation, see Sec. \ref{sec:data} for details. We leave these parameters free in the fitting analyses. The prior allows us to make sure that at $z>z_{\rm max}$ the shape of the Hubble function does not depart from the standard one, keeping also the physics at recombination untouched. The uncertainties of $(\tilde{\Omega}_m,\tilde{H}_0)$ have an impact on the observables at $z>z_{\rm max}$, and also propagate to the uncertainties of the parameters entering the Hubble function at smaller redshifts. This is why it is important to consider them. The radiation parameter appearing in the last row of Eq. \eqref{eq:H} reads $\tilde{\Omega}_r=4.18343\cdot 10^{-5}/\tilde{h}^2$, with $\tilde{h}=\tilde{H}_0/(100$ km/s/Mpc). The numerical coefficient is fixed by the current CMB temperature \cite{Fixsen:2009ug}, assuming for the sake of simplicity three relativistic neutrino species.

\begin{table*}
    \centering
    \begin{tabular}{ccccc}
        \hline
        Survey & $z$ & Observable & Measurement & References \\
        \hline
        \hline
        6dFGS+SDSS MGS & 0.122 & $D_V(r_{d}^{fid}/r_d)$  & $539\pm17$ [Mpc] & \cite{Carter_2018} \\
        \hline
         WiggleZ & 0.44 & $D_V(r_{d}^{fid}/r_d)$  & $1716.4\pm83.1$ [Mpc] & \cite{Kazin:2014qga}\\
         & 0.60 & $D_V(r_{d}^{fid}/r_d)$  & $2220.8\pm100.6$ [Mpc] & \\
         & 0.73 & $D_V(r_{d}^{fid}/r_d)$  & $2516.1\pm86.1$ [Mpc] & \\
         \hline
        BOSS DR12 & 0.32 & $r_{d}H/(10^{3} km/s)$ & $11.549\pm0.385$ & \cite{Gil_Mar_n_2016}\\
         & & $D_A/r_d$ & $6.5986\pm0.1337$ &\\
         & 0.57 & $r_{d}H/(10^{3} km/s)$ & $14.021\pm0.225$ & \\
         & & $D_A/r_d$ & $9.389\pm0.103$ &\\
         \hline
         DES Y3 & 0.835 & $D_M/r_d$ & $18.92\pm0.51$ & \cite{DES:2021esc}\\
         \hline
         Quasars eBOSS DR16 & 1.48 & $D_M/r_d$ & $30.21 \pm 0.79$ & \cite{Hou:2020rse}\\
          &  & $c/(Hr_d)$ & $13.23 \pm 0.47$ & 
         \\ \hline
         Ly$\alpha$-Forests eBOSS DR16 & 2.334  & $D_M/r_d$ & $37.5^{+1.2}_{-1.1}$ & \cite{duMasdesBourboux:2020pck}\\
          &  & $c/(H r_d)$ & $8.99^{+0.20}_{-0.19}$ &
        \\
        \hline
    \end{tabular} \caption{List with the 13 anisotropic BAO data points used in this work. The fiducial values of the comoving sound horizon appearing in the third column are $r_{d}^{fid} = 147.5$ Mpc for \cite{Carter_2018} and $r_{d}^{fid} = 148.6$ Mpc for \cite{Kazin:2014qga}. $D_M(z)=(1+z)D_A(z)$ is the comoving angular diameter distance and $D_V(z)=[D_M^2(z)cz/H(z)]^{1/3}$ is the so-called dilation scale. We have duly taken into account the existing internal correlations between the data points of WiggleZ, BOSS DR12, and QSOs and Ly$\alpha$ eBOSS DR16. See the quoted references for details.}\label{tab:BAO_data}
\end{table*}

We impose the following six constraints, which allow us to compute the parameters $\{a,b,c,d,e,f\}$ entering the functions $\delta H_1(z)$ and $\delta H_2(z)$,

\begin{equation}\label{eq:cond1}
\delta H_1(z=0)= H_0-\bar{H}_0\equiv\delta H_0 \,,
\end{equation}
\begin{equation}\label{eq:cond2}
  \delta H_1(z_p)=\delta H_2(z_p)\equiv \delta H_p \,,
\end{equation}

\begin{equation}\label{eq:cond3}
  \left.\frac{\partial \delta H_1}{\partial z}\right\vert_{z=z_p}=\left.\frac{\partial \delta H_2}{\partial z}\right\vert_{z=z_p}=0  \,,
\end{equation}

\begin{equation}\label{eq:cond4}
  \delta H_2(z_{\rm max}) = H(z_{\rm max})-\bar{H}(z_{\rm max})\equiv\delta H_{\rm max}\,,
\end{equation}
with $z_p$ the pivot redshift at which we have the extrema of $\delta H_1$ and $\delta H_2$. The two conditions in Eqs. \eqref{eq:cond2} and \eqref{eq:cond3} are obtained by demanding at $z_p$ the continuity of $H(z)$ and its derivative, respectively, whereas the condition \eqref{eq:cond4} enforces the continuity of the Hubble function at $z_{\rm max}$.

We consider the suite of parameters $\{\tilde{\Omega}_m,\tilde{H}_0,H_0,z_p,\delta H_p,z_{\rm max}\}$. The pair $\{\tilde{\Omega}_m,\tilde{H}_0\}$ and $H_0$ will be mainly controlled by the {\it Planck} and SH0ES priors, respectively. The triad $\{z_p,\delta H_p,z_{\rm max}\}$ is a priori more uncertain. However, we can already understand that if we set $z_{\rm max}$ to a value smaller than the smallest redshift in the BAO data set, $z_{\rm BAO,min}$, we will be unable to put strong constraints on the pair $\{z_p,\delta H_p\}$, since in this case there exists a full degeneracy between these two parameters. This degeneracy is essentially fixed by the CMB prior on the angular diameter distance to the last scattering surface, see Sec. \ref{sec:data}. Thus, for $z_{\rm max}\leq z_{\rm BAO,min}$ we can only constrain the direction of the degeneracy line. For $z_{\rm max}>z_{\rm BAO,min}$ the situation could be different, of course. We use uninformative flat priors for $\{z_p,\delta H_p\}$, much wider than the constraints we get from the fitting analyses.

The conditions \eqref{eq:cond1}-\eqref{eq:cond4} can be written in a very simple way,

\begin{equation}
\begin{pmatrix}
\delta H_0 \\
\delta H_p \\
0\\
\end{pmatrix}= \begin{pmatrix}
1 & 0 & 0\\
1 & z_p & z_p^2\\
0 & 1 & 2z_p
\end{pmatrix} 
\begin{pmatrix}
a \\
b \\
c\\
\end{pmatrix}
\end{equation}
and

\begin{equation}
\begin{pmatrix}
\delta H_{\rm max} \\
\delta H_p \\
0\\
\end{pmatrix}= \begin{pmatrix}
1 & z_{\rm max} & z_{\rm max}^2\\
1 & z_p & z_p^2\\
0 & 1 & 2z_p
\end{pmatrix} 
\begin{pmatrix}
d \\
e \\
f\\
\end{pmatrix}\,,
\end{equation}
so the constants $\{a,b,c,d,e,f\}$ needed to compute $H(z)$ can be obtained straightforwardly. We reconstruct the Hubble function by means of Monte Carlo analyses carried out with \texttt{Mathematica} \cite{Mathematica}, making use of Eq. \eqref{eq:H} and the Baseline data sets described in Sec. \ref{sec:data}\footnote{Our approach is similar to the one employed in \cite{Dutta:2018vmq}, but in this study we will show explicitly the need of an evolving $M(z)$ on top of the new physics at cosmological level. In addition, it is very important to notice that the authors of \cite{Dutta:2018vmq} did not employ a CMB prior on $D_A(z_*)$. They made use of a prior on $H(z_i)$ at several high redshifts ($z>4$) obtained with the {\it Planck}/$\Lambda$CDM cosmology. This is insufficient if we want to study the Hubble tension in a robust and unbiased way, since by considering only the constraints on $H(z_i)$ at $z>4$ from {\it Planck} we obtain curves of $H(z)$ that do not respect in general the very tight constraint we have on $D_A(z_*)$ and, hence, spoil the good description of the CMB data.}. We employ the Metropolis-Hastings algorithm \cite{1953JChPh..21.1087M,Hastings:1970aa} and stop the routine when the Monte Carlo performs $2\cdot 10^4$ jumps in parameter space, which is more than sufficient to achieve convergence in all the runs carried out in this work, with values of the Gelman-Rubin convergence statistic \cite{R2:1992,R1:1997} $R-1<0.02$ for all the sampled parameters. We set $z_{\rm max}$ to different values in order to study its impact in our analyses, and sample the five parameters contained in the vector $\{\tilde{\Omega}_m,\tilde{H}_0,H_0,z_p,\delta H_p\}$.   
Using the resulting $H(z)$ it is also possible to reconstruct in a trivial way higher order cosmographical functions as the deceleration parameter, which reads,

\begin{equation}\label{eq:q}
  q(z) = -1+\frac{(1+z)}{H(z)}\frac{dH}{dz}\,.  
\end{equation}
We will show results for $q(z)$ too in Sec. \ref{sec:results}. 


\subsection{Reconstruction of $M(z)$}\label{sec:recoM}

Let us consider the relation between the luminosity distance to a given object in a flat FLRW Universe,

\begin{equation}
D_L(z)=c(1+z)\int_0^z\frac{d\tilde{z}}{H(\tilde{z})}\,,
\end{equation} 
its apparent magnitude $m$ and its absolute magnitude $M$, which is given by

\begin{equation}\label{eq:M}
M = m-25-5\log_{10}\left(\frac{D_L}{1\,{\rm Mpc}}\right)\,.
\end{equation}
For standardizable objects, the standardized absolute magnitude is just a constant and, hence, does not depend on the position nor the redshift. This is what it is usually assumed for SNIa. Here, though, we abandon this assumption and consider that the usual standardization method of SNIa can still receive an unknown correction, making the absolute magnitude to evolve with the redshift. We use Gaussian Processes \cite{2006gpml.book.....R} to generate samples of $m(z)$ from the Pantheon+ SNIa compilation (see Sec. \ref{sec:data}), and combine them with our Markov chains of $H(z)$ to reconstruct $M(z)$. This allows us to assess whether the low-$z$ solutions to the $H_0$ tension require the evolution of $M$.   

\begin{table}
    \centering
   
    \label{tab:BAO2D_data}
    \begin{tabular}{ccccc}
        \hline
        $z$ & $\theta_{BAO}$ [deg] & $\sigma_{BAO}$ [deg] & References \\
        \hline
        \hline
        0.11 & 19.8 & 3.26 & \cite{deCarvalho:2021azj} \\ \hline
        0.235 & 9.06 & 0.23 & \cite{Alcaniz:2016ryy} \\
        0.365 & 6.33 & 0.22 \\ \hline
        0.45 & 4.77 & 0.17 & \cite{Carvalho:2015ica} \\
        0.47 & 5.02 & 0.25 \\
        0.49 & 4.99 & 0.21 \\
        0.51 & 4.81 & 0.17 \\
        0.53 & 4.29 & 0.30 \\
        0.55 & 4.25 & 0.25 \\ \hline
        0.57 & 4.59 & 0.36 & \cite{Carvalho:2017tuu} \\
        0.59 & 4.39 & 0.33 \\
        0.61 & 3.85 & 0.31 \\
        0.63 & 3.90 & 0.43 \\
        0.65 & 3.55 & 0.16 \\ \hline
        2.225 & 1.77 & 0.31 & \cite{deCarvalho:2017xye}      
        
        \\
        \hline
    \end{tabular} \caption{List with the 15 2D BAO data points used in this work, with $\theta_{\rm BAO}(z)\,[{\rm rad}]=r_d/[(1+z)D_A(z)]$. We employ a diagonal covariance matrix. See the quoted references for details.}\label{tab:BAO_data2D}
\end{table}

For the reconstruction of $m(z)$ we use the public package {\it Gaussian
Processes in Python} (GaPP)\footnote{https://github.com/carlosandrepaes/GaPP} \cite{Seikel:2012uu}. In particular, we use the Matérn 32 kernel and the optimization of its hyperparameters. This procedure has been already tested and employed in \cite{Favale:2023lnp}.

We present the results obtained using anisotropic and angular BAO data in Secs. \ref{sec:IDLanisotropic} and \ref{sec:IDLangular}, respectively.

 \begin{table}
    \centering
    
    \begin{tabular}{cccc}
        \hline
        $z$ & $H(z)$ [Km/s/Mpc] & References \\
        \hline
        \hline
        0.07 & 69.0$\pm$19.6 & \cite{Zhang:2012mp} \\ 
        0.09 & 69.0$\pm$12.0 & \cite{Jimenez:2003iv} \\
        0.12 & 68.6$\pm$26.2 & \cite{Zhang:2012mp} \\
        0.17 & 83.0$\pm$8.0 & \cite{Simon:2004tf} \\
        0.1791 & 78.0$\pm$6.2 & \cite{moresco2012improved} \\
        0.1993 & 78.0$\pm$6.9 & \cite{moresco2012improved} \\
        0.2 & 72.9$\pm$29.6 & \cite{Zhang:2012mp} \\
        0.27 & 77.0$\pm$14.0 & \cite{Simon:2004tf}  \\
        0.28 & 88.8$\pm$36.6 &\cite{Zhang:2012mp} \\
        0.3519 & 85.5$\pm$15.7 & \cite{moresco2012improved} \\
        0.3802 & 86.2$\pm$14.6 & \cite{Moresco:2016mzx} \\
        0.4 & 95.0$\pm$17.0 & \cite{Simon:2004tf}  \\
        0.4004 & 79.9$\pm$11.4 & \cite{Moresco:2016mzx} \\
        0.4247 & 90.4$\pm$12.8 & \cite{Moresco:2016mzx} \\
        0.4497 & 96.3$\pm$14.4 & \cite{Moresco:2016mzx} \\
        0.47 & 89.0$\pm$49.6 & \cite{Ratsimbazafy:2017vga} \\
        0.4783 & 83.8$\pm$10.2 & \cite{Moresco:2016mzx} \\
        0.48 & 97.0$\pm$62.0 & \cite{Stern:2009ep} \\
        0.5929 & 107.0$\pm$15.5 & \cite{moresco2012improved} \\
        0.6797 & 95.0$\pm$10.5 & \cite{moresco2012improved} \\
        0.75 & 98.8$\pm$33.6 & \cite{Borghi:2021rft} \\
        0.7812 & 96.5$\pm$12.5 & \cite{moresco2012improved} \\
        0.8754 & 124.5$\pm$17.4 & \cite{moresco2012improved} \\
        0.88 & 90.0$\pm$40.0 & \cite{Stern:2009ep} \\
        0.9 & 117.0$\pm$23.0 & \cite{Simon:2004tf}  \\
        1.037 & 133.5$\pm$17.6 & \cite{moresco2012improved} \\
        1.26 & $135.0\pm65.0$ & \cite{Tomasetti:2023kek} \\
        1.3 & 168.0$\pm$17.0 & \cite{Simon:2004tf}  \\
        1.363 & 160.0$\pm$33.8 & \cite{Moresco:2015cya} \\
        1.43 & 177.0$\pm$18.0 & \cite{Simon:2004tf}  \\
        1.53 & 140.0$\pm$14.0 & \cite{Simon:2004tf}  \\
        1.75 & 202.0$\pm$40.0 & \cite{Simon:2004tf}  \\
        1.965 & 186.5$\pm$50.6 & \cite{Moresco:2015cya} \\
        \hline
    \end{tabular}\caption{List with the 33 CCH data points on $H(z)$ used in this work, obtained from the references quoted in the last column. In the case of Refs. \cite{moresco2012improved, Moresco:2016mzx}, the central values of $H(z)$ are computed by performing the arithmetic mean of the measurements obtained with the BC03 \cite{Bruzual:2003tq} and MaStro \cite{Maraston:2011sq} stellar population synthesis models\footnote{This is the origin of the differences with Table 1.1 of \cite{Moresco:2023zys}, in which the author reported the BC03 values, which are in most cases larger than those obtained with MaStro.}. The covariance matrix is computed using the method presented in \cite{Moresco:2020fbm}, which incorporates both the statistical and systematic errors. See the quoted references for details.}\label{tab:CCH}
\end{table}                      


\subsection{Assessing the fulfillment of the weak energy condition with the aid of CCH}\label{sec:WECmethod}

The weak energy condition is fulfilled if the null energy condition

\begin{equation}\label{eq:NEC}
T_{\mu\nu}k^\mu k^\nu\geq 0
\end{equation}
is satisfied for any null vector ($k^\mu k_\mu=0$) together with

\begin{equation}\label{eq:WEC}
T_{\mu\nu}u^\mu u^\nu\geq 0\,,
\end{equation}
where $T_{\mu\nu}$ is the energy-momentum tensor and $u^\mu$ is a time-like four-vector. For a perfect fluid with density $\rho$ and pressure $p$ in a FLRW universe, Eqs. \eqref{eq:WEC} and \eqref{eq:NEC} translate into the two conditions

\begin{equation}\label{eq:WEC2}
\rho\geq 0\qquad {\rm and}\qquad \frac{d\rho}{dz}\geq 0\,,
\end{equation}
which essentially stand for the positivity of the energy density and its constant or decaying nature. They are satisfied by all the fluids considered in the $\Lambda$CDM model, so the violation of the WEC would automatically imply the existence of physics beyond the standard model\footnote{The strong energy condition reads, instead: $\rho+p\geq 0$ and $\rho+3p\geq 0$. The cosmological constant satisfies the weak energy condition, but not the strong one.}. Here we are interested in testing the fulfillment of the WEC by the effective dark energy fluid in charge of the accelerated expansion of the universe, considering that it is covariantly self-conserved and, hence, that matter is diluted according to the usual law\footnote{Scenarios with a coupling with matter are also interesting, but the analysis would depend on the choice of the source vector, which controls the transfer of energy and momentum between the dark components.}, 

\begin{equation}
\rho_m(z)=\rho_m^0(1+z)^3\,.
\end{equation}
 It was shown in \cite{Sen:2007ep} that in this case the Friedmann equation together with the second inequality of Eq. \eqref{eq:WEC2} lead to 

\begin{equation}\label{eq:wec}
\Omega_m\leq \frac{E^2(z)-1}{(1+z)^3-1}\,,
\end{equation}                   
with $E(z)=H(z)/H_0$ the normalized Hubble rate. We denote the right-hand side of this inequality as $\Omega_m^{\rm max}$. It is an upper bound that cannot be surpassed if the effective dark energy fluid is not phantom. As already shown in \cite{Sen:2007ep}, one can use cosmic chronometers in combination with a prior on $H_0$ to obtain as many estimates of $\Omega_m^{\rm max}$ as data points we have on CCH\footnote{Similar methods, as the $Om$ and $Omh^2$ diagnostics, have also proved useful to test the $\Lambda$CDM model \cite{Shafieloo:2012rs,Sahni:2014ooa}.}. One can sample the Gaussian distributions of $H_0$ and the CCH data to obtain a chain with $\Omega_m^{\rm max}(z_i)$.

We aim to improve the analysis of \cite{Sen:2007ep} in several ways, namely: (i) we take advantage of the larger sample of CCH measurements (we have now 33 data points instead of 9), see Sec. \ref{sec:data}; (ii) we also employ their corresponding covariance matrix; and (iii) we apply an advanced method to get a single representative value of $\Omega_m^{\rm max}$, duly accounting for the correlations and assessing the impact of non-Gaussian features in the multivariate distribution.  

The latter is done by means of the so-called Edgeworth expansion. It allows us to compute an analytical approximation of the underlying (exact) distribution,

\begin{equation}\label{eq:Edgeworth}
\begin{split}
f(\vec{x})= G(\vec{x},\lambda)[1+&\frac{1}{6}k^{ijk}h_{ijk}(\vec{x},\lambda)\\
&+\frac{1}{24}k^{ijkl}h_{ijkl}(\vec{x},\lambda)+...]\,,
\end{split}
\end{equation}
see \cite{Amendola:1996xwd} and references therein. Here $x^i=d^i-\mu^i$, with $d^i=\Omega_m^{\rm max}(z_i)$ and $\vec{\mu}$ the mean vector, i.e. $\mu^i=<\Omega_m^{\rm max}(z_i)>$. $\lambda=C^{-1}$ is the inverse of the covariance matrix, with elements $C^{ij}=<x^i x^j>$. $G(\vec{x},\lambda)$ is the multivariate Gaussian distribution built from that mean and covariance matrix, and $k^{ijk}=<x^ix^jx^k>$ and $k^{ijkl}=<x^ix^jx^kx^l>-C^{ij}C^{kl}-C^{ik}C^{jl}-C^{il}C^{jk}$ are the elements of the higher-order cumulant matrices, called skewness and kurtosis matrices, respectively. On the other hand, 
             
\begin{equation}
h_{ij...}(\vec{x},\lambda)=(-1)^r G^{-1}(\vec{x},\lambda)\partial_{ij...} G(\vec{x},\lambda)\,,
\end{equation}
are the Hermite tensors of order $r$, with $r$ the number of indices. The Hermite tensors of order 3 and  4 appearing in Eq. \eqref{eq:Edgeworth} read, respectively, 

\begin{equation}
h_{ijk}(\vec{x})=\lambda_{in}\lambda_{jt}\lambda_{kl}x^nx^tx^l-(\lambda_{ij}\lambda_{kt}+\lambda_{ik}\lambda_{jt}+\lambda_{jk}\lambda_{it})x^t\,,
\end{equation}
\begin{equation}
\begin{split}
h_{lijk}(\vec{x})=&\lambda_{ln}x^nh_{ijk}(\vec{x})+\lambda_{ij}\lambda_{kl}+\lambda_{ik}\lambda_{jl}+\lambda_{jk}\lambda_{il}\\
&-(\lambda_{il}\lambda_{jt}\lambda_{kn}+\lambda_{in}\lambda_{jl}\lambda_{kt}+\lambda_{in}\lambda_{jt}\lambda_{kl})x^nx^t\,.
\end{split}
\end{equation}
In this calculation we have made use of the fact that $\partial_i G = -G \lambda_{ij} x^j$ and of Einstein's summation convention. All these objects can be directly computed from the chain of $\Omega_m^{\rm max}(z_i)$, with $i$ the index that runs over the CCH data points. Once we have Eq. \eqref{eq:Edgeworth}, we can sample it treating it as a one-dimensional distribution for $\Omega_m^{\rm max}$ (instead of a multivariate distribution for the array $\{\Omega_m^{\rm max}(z_i)\}$). If the non-Gaussian features are negligible, then it reduces of course to a Gaussian with the following weighted mean and variance,

\begin{figure*}    \centering
\includegraphics[scale=0.7]{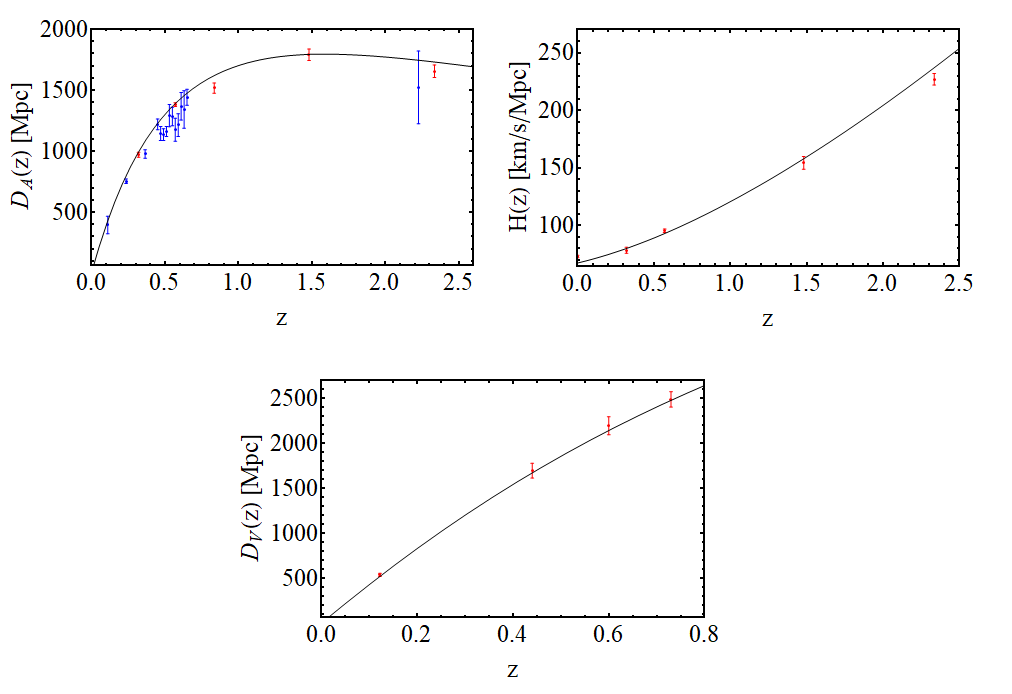}    \caption{Calibrated 3D and 2D BAO data, in red and blue, respectively. The calibration is carried out with the {\it Planck}/$\Lambda$CDM value of the sound horizon $r_d^{P18}$, see Sec. \ref{sec:data}. We also plot (in black) the curves of the various observables computed with the best-fit {\it Planck}/$\Lambda$CDM cosmology. See the comments in the first paragraph of Sec. \ref{sec:results}.}\label{fig:BAOdata2}
\end{figure*}

\begin{equation}\label{eq:Gapprox}
\bar{\Omega}_m^{\rm max}=\frac{\sum\limits_{i,j=1}^{33}\mu^i\lambda_{ij}}{\sum\limits_{i,j=1}^{33}\lambda_{ij}}\qquad ;\qquad \sigma^2=\frac{1}{\sum\limits_{i,j=1}^{33}\lambda_{ij}}\,.
\end{equation} 
Otherwise, Eq. \eqref{eq:Edgeworth} introduces some corrections in the  computation of the central value and confidence intervals of $\Omega_m^{\rm max}$. 

Another possibility to test the WEC is to consider the following relation,

\begin{equation}\label{eq:ratioDE}
\frac{\rho_{\rm de}(z)}{\rho_{\rm de}^0}=\frac{E^2(z)-\Omega_m(1+z)^3}{1-\Omega_m}\,,
\end{equation}
which is directly obtained from the Friedmann equation. Sampling $H(z)$ from the  CCH data and combining this information with a prior on $H_0$ and $\omega_m$ that lets us compute $\Omega_m$, it is possible to constrain Eq. \eqref{eq:ratioDE} at the redshifts of the CCH. Negative values of this quantity hint at a violation of the first WEC of Eq. \eqref{eq:WEC2}. Instead, $0<\rho_{\rm de}(z)/\rho_{\rm de}^0<1$ means that the second condition is not fulfilled.

We provide the results of these analyses in Sec. \ref{sec:WEC}.


\subsection{The data}\label{sec:data}

In this work we make use of the following data sets:

\begin{itemize}
\item The SH0ES prior on the Hubble parameter, $H_0^{R22}$ \cite{Riess:2021jrx}.

\item The anisotropic (3D) BAO data listed in Table \ref{tab:BAO_data}. 

\item The transversal (angular, 2D) BAO data listed in Table \ref{tab:BAO_data2D}. Angular BAO might be less model-dependent than 3D BAO, but have larger error bars \cite{deCarvalho:2021azj,Nunes:2020hzy}.

\item The {\it Planck} 2018 CMB TT,TE,EE+lowE+lensing $\Lambda$CDM Gaussian priors on the quantities $\{D_A(z_*),r_d,\tilde{\Omega}_m,\tilde{H}_0\}$, including the corresponding covariance matrix $C$. This information can be obtained from the {\it Planck} Legacy Archive\footnote{http://pla.esac.esa.int/pla/$\#$home}. They read, respectively,

 \begin{equation}\label{eq:priorPlanck}
\begin{pmatrix}
D_A(z_*)\,[{\rm Gpc}]\\
r_d\,[{\rm Mpc}]\\
\tilde{\Omega}_m\\
\tilde{H}_0\,[{\rm Mpc/km/s}]
\end{pmatrix}=\begin{pmatrix}
13.872\\
147.09\\
0.3153\\
67.36
\end{pmatrix}\,,
\end{equation}

\begin{equation}
C=\begin{pmatrix}
6.11\cdot 10^{-4} & 6.42\cdot 10^{-3} & -1.49\cdot 10^{-4} & 9.84\cdot 10^{-3} \\
- & 7.04\cdot 10^{-2} & -1.51\cdot 10^{-3} & 9.85\cdot 10^{-2}\\
- & - & 5.44\cdot 10^{-5} & -3.94\cdot 10^{-3} \\
- & - & - & 2.90\cdot 10^{-1} 
\end{pmatrix}\,,
\end{equation}
where the units of the elements of $C$ are set by those employed in Eq. \eqref{eq:priorPlanck}, and we omit the lower half of the covariance matrix, since it is obviously symmetric. The use of this prior is fully justified by our assumption of standard physics before recombination. By generating samples of these parameters and the BAO data out of the respective multivariate Gaussian distributions we can obtain a joint data vector which incorporates the BAO angular diameter distances, dilation scales, $H(z_i)$, together with $\{D_A(z_*),\tilde{\Omega}_m,\tilde{H}_0\}$, and their joint covariance matrix. The prior from SH0ES can be easily added, considering 0 correlation with the other parameters, since it is independent. We build the resulting vector of data and covariance matrix before performing the fitting analysis, of course. This will constitute our Baseline\_3D or Baseline\_2D data sets, depending on the BAO data set that we consider. They are 17- and 19-dimensional, respectively.

\begin{figure*}[t!]
\includegraphics[scale=0.45]{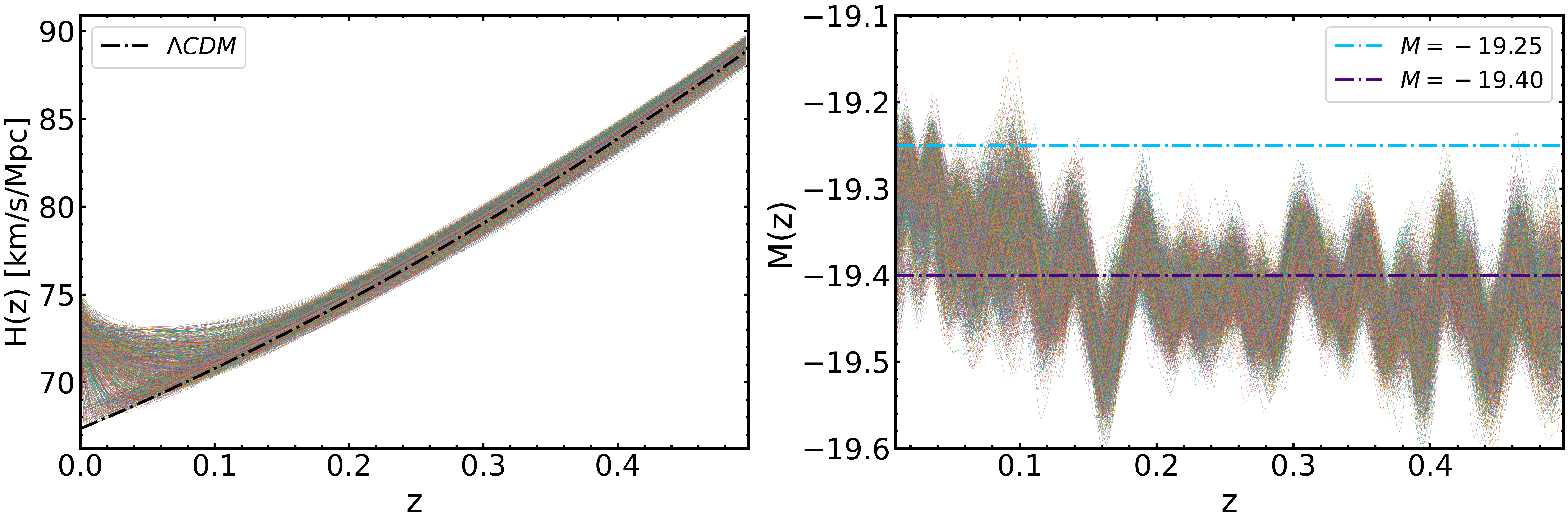} \caption{On the left, curves of $H(z)$ obtained with the Baseline\_3D data set, using Eq. \eqref{eq:H} and fixing $z_{\rm max}=1$, cf. Sec. \ref{sec:Hmethod}. We show the 68\% curves with smallest $\chi^2$. On the right, the corresponding absolute magnitude of SNIa, $M(z)$, see Sec. \ref{sec:recoM}. We zoom in the region $z<0.5$ to better appreciate the increase of $H(z)$ and $M(z)$ at small redshifts. At $z\gtrsim 0.2$ $H(z)\simeq H_\Lambda(z)$ (see Eq. \ref{eq:HLambda}) and $M(z)$ becomes compatible with the value obtained with the inverse distance ladder assuming the $\Lambda$CDM, $M\sim -19.4$ mag. }\label{fig:H_M_3D_joined}
\end{figure*}

\item The data on cosmic chronometers and covariance matrix used in \cite{Favale:2023lnp}. We add the new data point $H(z=1.26) = (135 \pm  65)$ km/s/Mpc \cite{Tomasetti:2023kek}, for completeness. See Table \ref{tab:CCH} and references therein. In some of our analyses we have tried the combination Baseline+CCH. In Sec. \ref{sec:WEC} we employ CCH together with the SH0ES or {\it Planck} priors on the Hubble parameter ($H_0^{R22}$ and $H_0^{P18}$, respectively) to study the weak energy condition applying the methods described in Sec. \ref{sec:WECmethod}. 

\item The SNIa data from the Pantheon+ compilation \cite{Scolnic:2021amr}. As explained in Sec. \ref{sec:recoM}, we do not employ them in the fitting analyses, but to reconstruct $M(z)$ with the help of the shapes of $H(z)$ obtained in the Monte Carlo runs.

\end{itemize}


\section{Results and discussion}\label{sec:results}

In Fig. \ref{fig:BAOdata2} we plot the cosmic distances and values of $H(z)$ obtained from the angular and anisotropic BAO data after their calibration with $r_d^{P18}$, together with the best-fit {\it Planck}$/\Lambda$CDM curves of the corresponding cosmological functions. The agreement between the latter and the vast majority of the 3D BAO data is clear, and this is why in order to accommodate the SH0ES measurement one has to modify the model at redshifts below $\sim 0.1$, which is the minimum BAO redshift in the 3D data set, cf. Table \ref{tab:BAO_data2D}\footnote{There are only mild tensions with the DES Y3 data point at $z=0.835$ and the Ly$\alpha$-Forest data at $z=2.334$ from eBOSS DR16, but they are much less significant than the Hubble tension.}. Conversely, there is an obvious tension between the best-fit {\it Planck}$/\Lambda$CDM cosmology and the 2D BAO data in the redshift range $0.4\lesssim z\lesssim 0.7$, since all the blue points of the upper-left plot in that redshift range fall below the black curve. These 2D BAO angular diameter distances are smaller than preferred by the standard model. This implies that 2D BAO data prefers larger values of the Hubble function, more in agreement with SH0ES \cite{Camarena:2019rmj}. Hence, in order to reestablish the concordance in the context of the distance ladder built with {\it Planck} and the 2D BAO data we need to modify the shape of $H(z)$ at much higher redshifts too, $z\sim 1$. We devote Secs. \ref{sec:IDLanisotropic} and \ref{sec:IDLangular} to show all this explicitly, together with the shapes of $M(z)$ needed to keep the consistency with the SNIa data.

\subsection{Analysis with Baseline\_3D}\label{sec:IDLanisotropic}

We present in the left plot of Fig. \ref{fig:H_M_3D_joined} the reconstructed curves of $H(z)$ obtained from the fit to the Baseline\_3D data set, following the method explained in Sec. \ref{sec:Hmethod}. More concretely, we show only the $68\%$ curves with lowest $\chi^2$. The right plot of Fig. \ref{fig:H_M_3D_joined}, instead, contains the associated curves of $M(z)$, obtained as described in Sec. \ref{sec:recoM}\footnote{We do not show in this work the mean and the corresponding confidence intervals of the various reconstructed functions, since they complicate the interpretation of the results due to the non-negligible impact of volume effects.}.

\begin{figure}[t!]  
\includegraphics[scale=0.9]{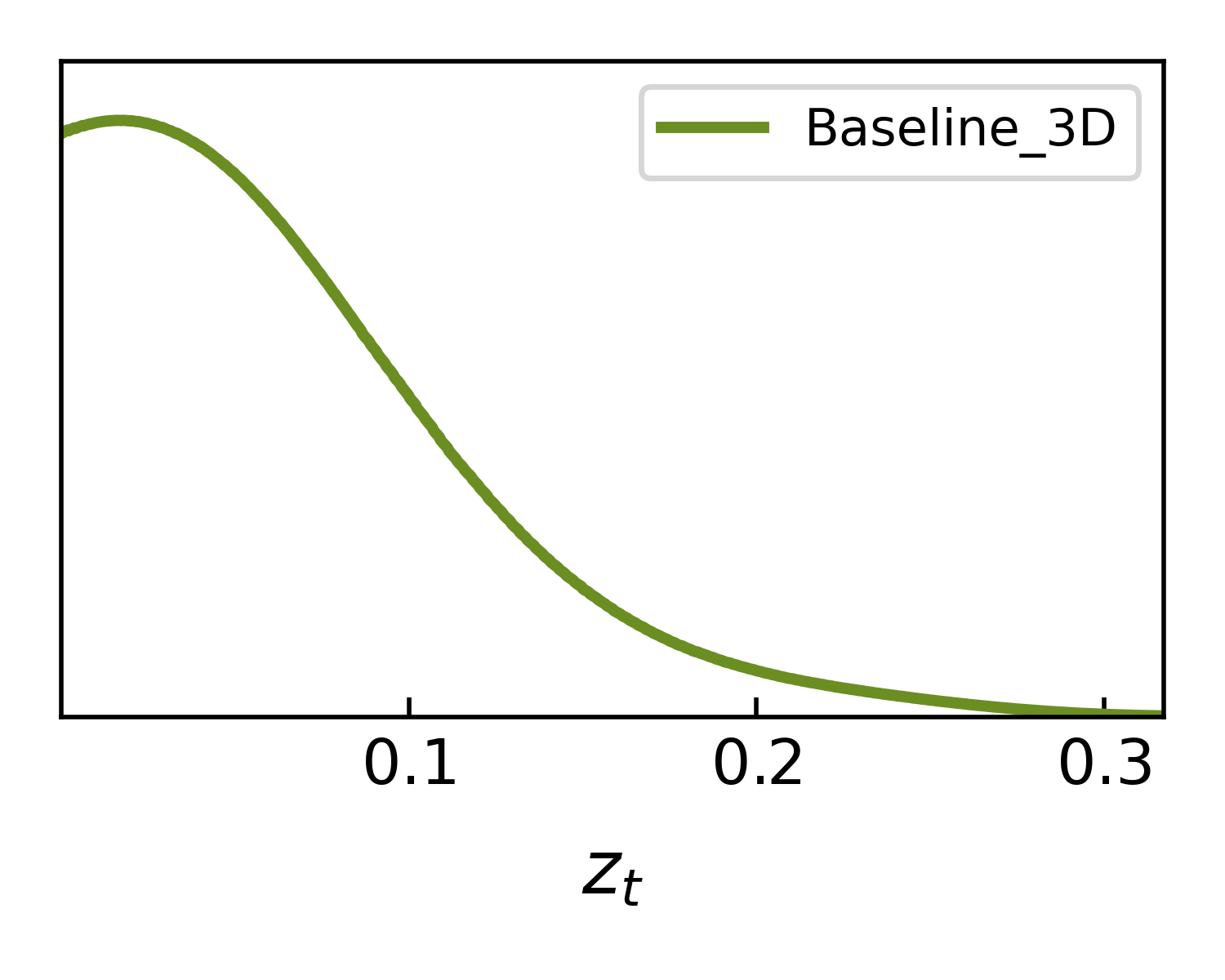}
\includegraphics[scale=0.9]{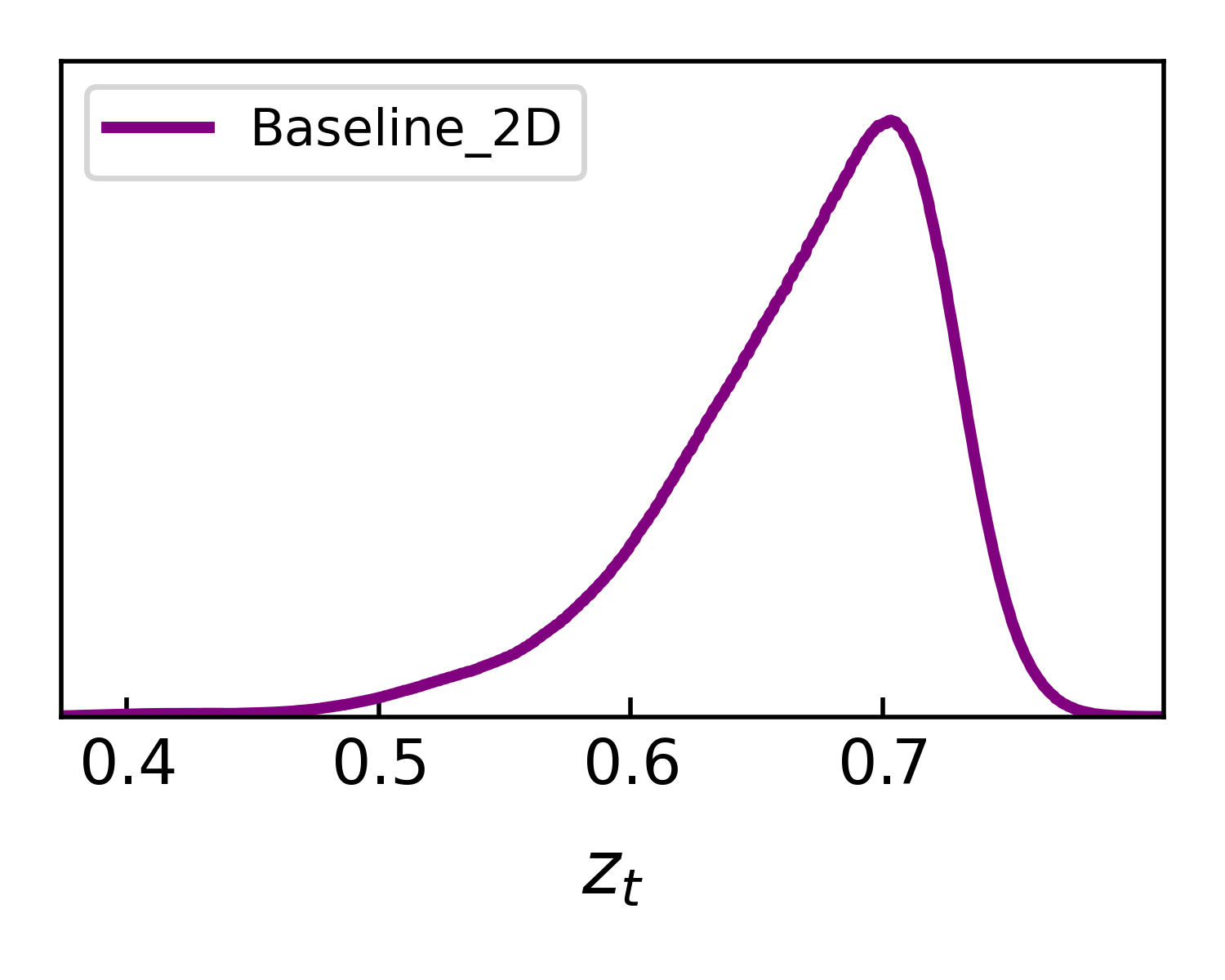} \caption{Posterior distributions of the transition redshift $z_t$ obtained with the Baseline\_3D and Baseline\_2D data sets from the corresponding Monte Carlo analyses.}\label{fig:post_zt}
\end{figure}

In this analysis we have fixed $z_{\rm max}=1$, but we have checked that values of $z_{\rm max}=0.5,2,4$ do not introduce any significant change in the shape of the reconstructed function and the minimum values of the $\chi^2$. Therefore, the results obtained with the baseline\_3D data set are stable and basically insensitive to the parameter $z_{\rm max}$. We find in all cases a phantom-like increase of $H(z)$ at $z\lesssim 0.2$ with respect to the best-fit {\it Planck}/$\Lambda$CDM model, and slightly smaller values of $H(z)$ at higher redshifts. This is consistent with the results reported in \cite{Alestas:2020zol}. This upper bound is close to the minimum redshift of the 3D BAO data, as expected from Fig. \ref{fig:BAOdata2}. In contrast, there is no lower bound on the redshift at which the non-standard growth of $H(z)$ starts. This means that the Baseline\_3D data set is not capable of arbitrating between an ultra-late-time transition at $z\lesssim 0.01$ and a late-time transition at $0.01\lesssim z\lesssim 0.2$, corresponding to the scenarios 2 and 3 discussed in the Introduction, respectively. We denote this (transition) redshift as $z_t$, and define it as the redshift at which $H(z_t)=1.005H_\Lambda(z_t)$, see its posterior distribution in the upper plot of Fig. \ref{fig:post_zt}. The smaller is $z_t$ the faster the increase of $H(z)$ and $M(z)$ in the transition, of course. 

\begin{figure}[t!]    
\includegraphics[width=\columnwidth]{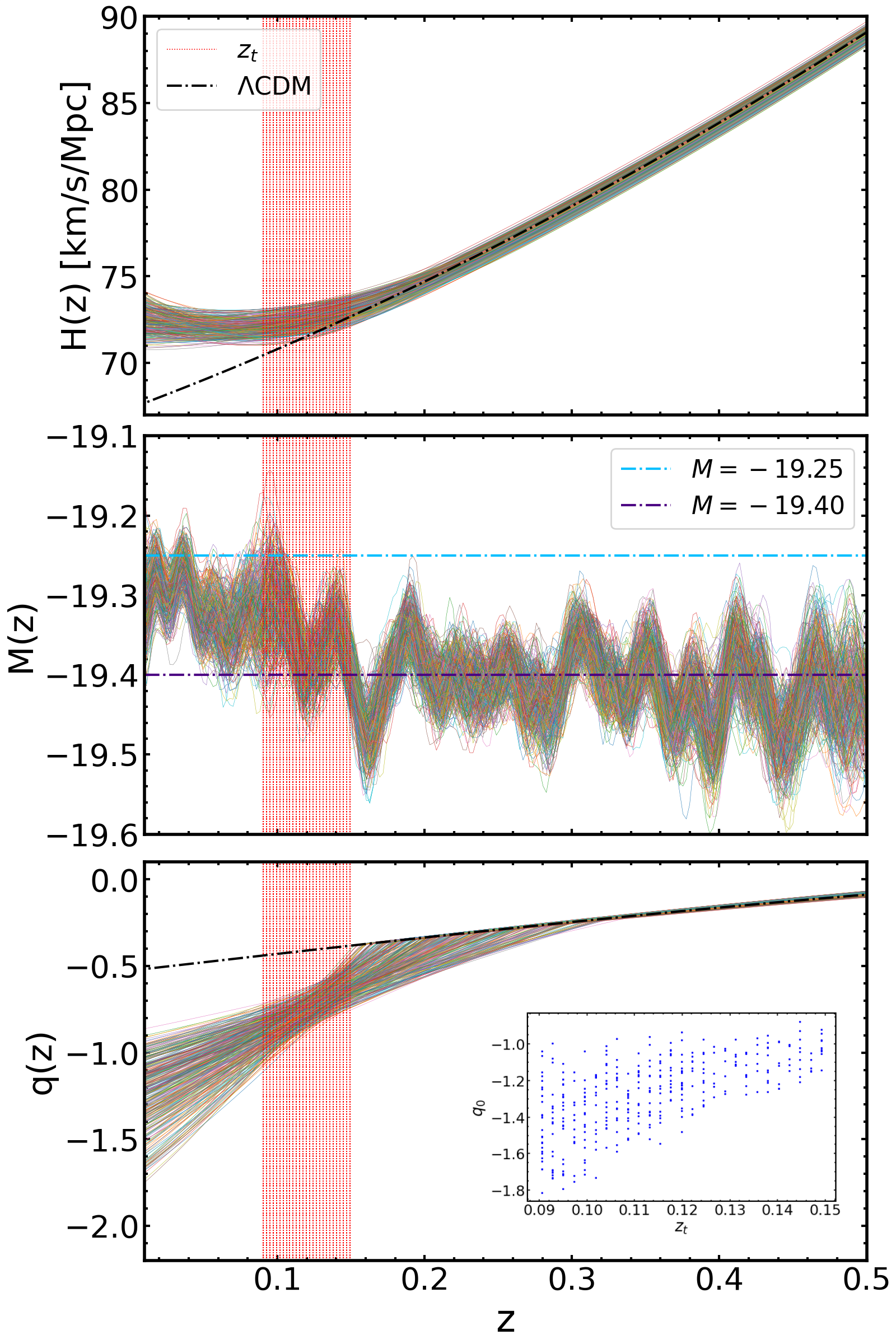}    \caption{Same as in Fig. \ref{fig:H_M_3D_joined}, but only for the curves with $0.09<z_t<0.15$. The vertical red band indicates the range of values of $z_t$ covered. In the middle plot, we include the constant values $M^{R22}$ (in cyan) and $M=-19.40$ (in purple), the latter being close to the $\Lambda$CDM best-fit value obtained from a CMB+BAO+SNIa analysis (see e.g. \cite{Gomez-Valent:2022hkb}). In the bottom plot, we present the corresponding shapes of the deceleration parameter $q(z)$, Eq. \eqref{eq:q}, and include an inner plot with the positive correlation between its value at $z=0$, $q_0$, and $z_t$. }\label{fig:transition0.1}
\end{figure}

Although there is a clear trend towards the value of $M$ measured by SH0ES at $z\sim 0$, $M^{R22}$, it is not easy to grasp the details of the transition for a fixed $z_t$ due to the large number of curves contained in the plots of Fig. \ref{fig:H_M_3D_joined}. This is why we choose to show in Fig. \ref{fig:transition0.1} only those curves with $0.09<z_t< 0.15$, as an illustrative example.

Also, we study an additional consequence of the fast increase of the Hubble function at those very low redshifts: a natural drop in the deceleration parameter, given by Eq. \eqref{eq:q}. Its shape is shown in the bottom plot of Fig. \ref{fig:transition0.1}. The values of $q(z=0)\equiv q_0$ are typically 2 or 3 times smaller than the one preferred by $\Lambda$CDM, $q_0\approx -0.55$, and can be even larger for lower values of $z_t$\footnote{These small values of $q_0$ would be excluded if we considered a fixed $M$ \cite{Gomez-Valent:2018gvm}.}. We also observe a positive correlation between the transition redshift and $q_0$, as expected. We show this in the inner plot. The smaller the value of $z_t$ the more accelerated has to be the Universe to transition to $H_0^{R22}$ and, hence, the more negative the value of $q_0$.

\begin{figure}[t!]  
\includegraphics[scale=0.7]{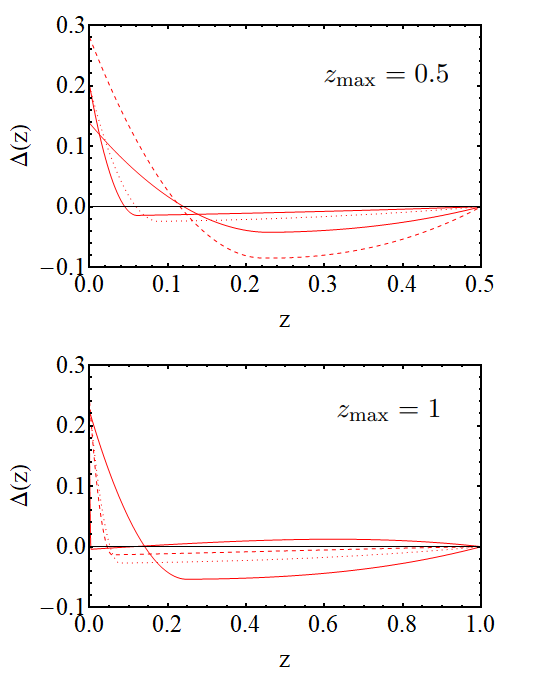}    \caption{Curves of $\Delta(z)$ (Eq. \eqref{eq:Delta}) obtained in the analysis of the baseline\_3D data set with $z_{\rm max}=0.5$ (upper plot) and $z_{\rm max}=1$ (lower plot). They have been selected from the group of 68\% curves with lowest $\chi^2$ in the Monte Carlo Markov chain and represent the typical behavior of $\Delta(z)$ required to solve the Hubble tension. The transition happens in both analyses at $z_t\lesssim 0.2$. See the main text for details.}  \label{fig:Delta3D}
\end{figure}

It is also a good exercise to study what is the typical shape of the density of a hypothetical effective dark energy fluid in these low-redshift solutions, provided that such a fluid is covariantly self-conserved. We define

\begin{equation}
\begin{split}
    H^2(z)&\equiv \frac{8\pi G}{3}[\tilde{\rho}_m^0(1+z)^3+\rho_{\rm de}(z)]\\
		&=\tilde{\Omega}_m\tilde{H}_0^2(1+z)^3+\frac{8\pi G}{3}\rho_{\rm de}(z)\,,
\end{split}
\end{equation}
with $H(z)$ given by Eq. \eqref{eq:H}, and also use

\begin{equation}
\begin{split}
    H_\Lambda^2(z)&=\frac{8\pi G}{3}[\tilde{\rho}_m^0(1+z)^3+\tilde{\rho}_{\Lambda}] \\
		&= \tilde{\Omega}_m\tilde{H}_0^2(1+z)^3+\frac{8\pi G}{3}\tilde{\rho}_{\Lambda}\,,
\end{split}
\end{equation}
where $H_\Lambda(z)$ takes the same form already assumed for $H(z)$ at $z>z_{\rm max}$ (cf. Eq. \eqref{eq:HLambda}) and $\tilde{\rho}_\Lambda$ is the energy density associated to the cosmological constant in the $\Lambda$CDM model. Using these two expressions we find the following relative difference between the effective dark energy density $\rho_{\rm de}(z)$ and $\tilde{\rho}_\Lambda$,

\begin{equation}\label{eq:Delta}
\Delta(z)\equiv \frac{\rho_{\rm de}(z)-\tilde{\rho}_{\Lambda}}{\tilde{\rho}_{\Lambda}}=\frac{H^2(z)-H_\Lambda^2(z)}{H_\Lambda^2(z)-\tilde{\Omega}_m\tilde{H}_0^2(1+z)^3}\,.
\end{equation}
In Fig. \ref{fig:Delta3D} we show some shapes of $\Delta(z)$ obtained with $z_{\rm max}=0.5$ and $z_{\rm max}=1$. They have been selected from the group of 68\% curves with lowest $\chi^2$ in the Monte Carlo Markov chains, and their form can be considered to be quite representative of the behavior that is required to explain the data and alleviate the Hubble tension. By construction of Eq. \eqref{eq:H}, we always find a crossing of the phantom divide, of course. Nevertheless, we want to remark that the steepness of the quintessence evolution between $z_p$ and $z_{\rm max}$ depends on both parameters. Larger values of $z_p$ and smaller values of $z_{\rm max}$ favor a transition from quintessence to phantom, but the Baseline\_3D data set cannot exclude a solution with an almost constant $\tilde{\rho}_{\rm de}$ before the transition for some combinations of these two parameters. What is always needed is a very fast phantom evolution in the last stages of the cosmic expansion, which is faster for smaller values of $z_t$. Our results resonate well with the conclusions of \cite{Alestas:2020zol,Heisenberg:2022gqk,Heisenberg:2022lob}, and also with \cite{DiValentino:2020naf,Adil:2023exv}. In addition, if we decrease $z_{\rm max}$ we can find more negative values of $\Delta$ at its minimum. 

\begin{figure}[t!]   
\includegraphics[width=\columnwidth]{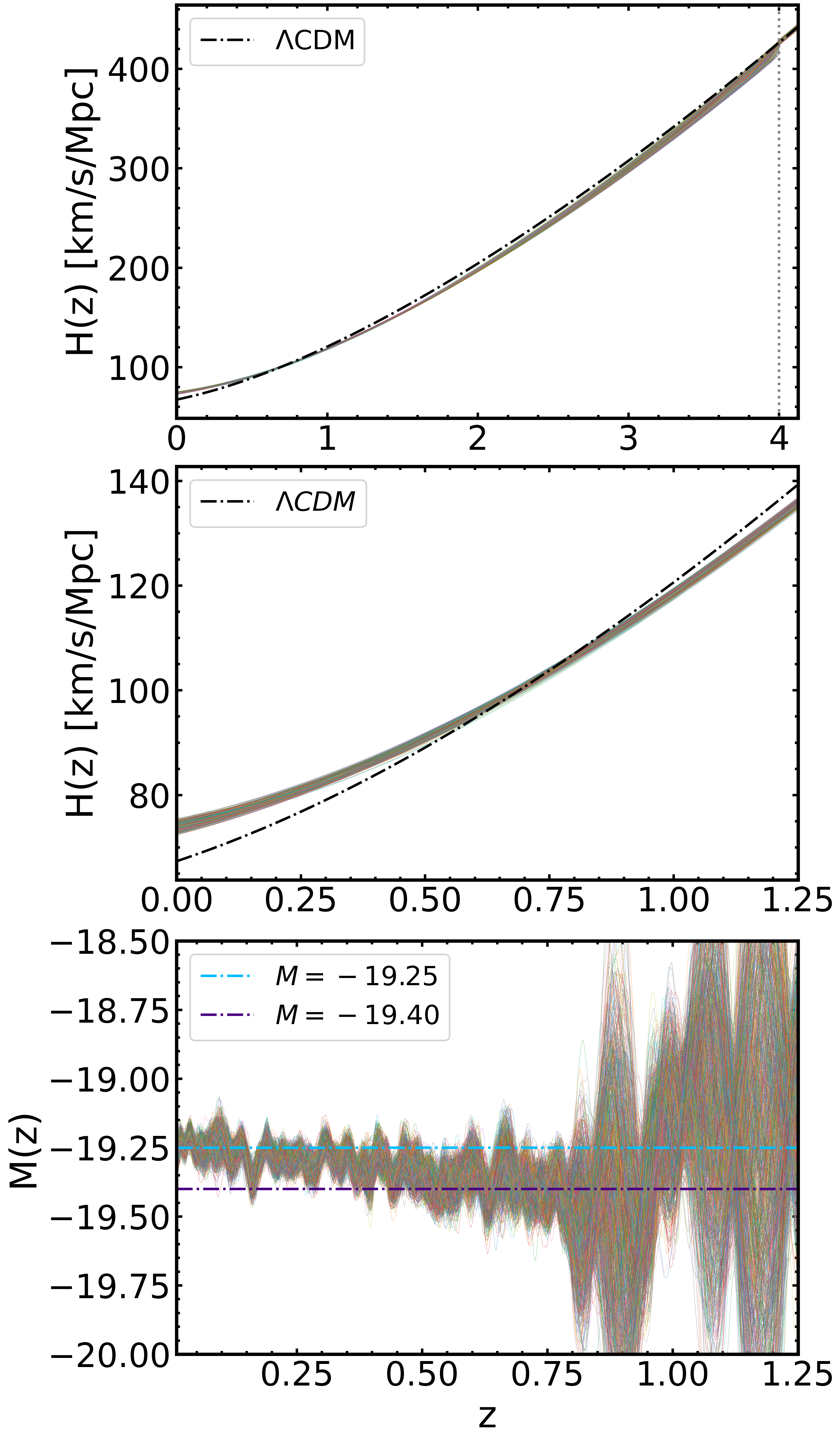}
\caption{Same as in Fig. \ref{fig:H_M_3D_joined}, but with the baseline\_2D data, using $z_{\rm max}=4$. The curves of $H(z)$ deviate from the {\it Planck}/$\Lambda$CDM prediction in a broader redshift range, with a transition from $H<H_{\Lambda{\rm CDM}}$ to $H>H_{\Lambda{\rm CDM}}$ happening now at $z_t\sim 0.5-0.8$ (see the zoomed-in middle plot). This is reflected also in the transition of $M(z)$ (bottom plot), which happens also at much higher $z$ than in the 3D BAO case, cf. again Fig. \ref{fig:H_M_3D_joined}.} \label{fig:H_M_2D_joined}
\end{figure}

In order to study the goodness of fit, we have computed the reduced $\chi^2$,

\begin{equation}\label{eq:reducedchi2}
\chi^2_{\rm red}=\chi^2_{\rm min}/(N-n)\,,
\end{equation}
with $N=17$ the number of data points and $n=5$ the number of fitting parameters, cf. Secs. \ref{sec:Hmethod} and \ref{sec:data}. We obtain $\chi^2_{\rm red}\sim 1.3$, slightly above 1. This can be explained essentially by the  $\sim 2\sigma$ tension between $\Lambda$CDM and the BAO data from DES and Ly$\alpha$ eBOSS DR16, and also by the fact that the effective number degrees of freedom can be significantly larger due to the existence of strong correlations between the parameters, see e.g. \cite{Andrae:2010gh}. If we remove the aforesaid data points, we get $\chi^2_{\rm red}\lesssim 1$. In any case, the value $\chi^2_{\rm red}\sim 1.3$ corresponds to a p-value $=0.21$ for $N-n=12$ degrees of freedom, hence we can conclude that our fitting function is performing well.

Finally, we have also checked that the inclusion of the CCH data does not alter our results in a significant way. We will explain why in Sec. \ref{sec:WEC}.   

In the context of concrete cosmological models, one can in principle study the evolution of perturbations to put tighter constraints on the class of viable ultra-low-redshift solutions to the $H_0$ tension. We expect the growth data to prefer an effective crossing of the phantom divide in order not to worsen the tension with galaxy clustering and weak lensing observations \cite{Heisenberg:2022gqk,Heisenberg:2022lob}\footnote{For details on this tension see e.g. \cite{Perivolaropoulos:2021jda,DiValentino:2020vvd}, and references therein.}. For the sake of generality, this study is left for future research, but we remark that a joint solution to the Hubble and growth tensions in the context of models with a transition in the redshift range $0.01\lesssim z\lesssim 0.2$ implies the existence of a late-time effective phantom regime accompanied by a fast increase of the absolute magnitude of SNIa, and most probably also a crossing of the phantom divide. A transition in the Hubble rate alone is certainly unable to address the $H_0$ tension according to the Baseline\_3D data set, see also \cite{Keeley:2022ojz}. These effects at very small redshifts would most probably introduce a new coincidence (or ``why now'') problem. The simultaneous increase of $H$ and $M$ could indicate a gravitational origin of these transitions and a hint of deviations from General Relativity. If, instead, the transition happens at $z\lesssim 0.01$ we retrieve the scenario 2 of the Introduction, and the possible solutions discussed therein.


\subsection{Analysis with Baseline\_2D}\label{sec:IDLangular}

\begin{figure}[t!]   
\includegraphics[scale=0.55]{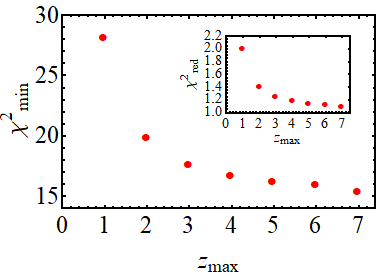}
\caption{Values of the minimum $\chi^2$ obtained in the fitting analyses with the baseline\_2D data set, as a function of $z_{\rm max}$. We also show the reduced $\chi^2$ (Eq. \eqref{eq:reducedchi2}) in the inner plot. The goodness of fit reaches a plateau at $z_{\rm max}\sim 4$. This explains why the transition redshift remains stable for $z_{\rm max}\gtrsim 4$, with $0.5\lesssim z_t\lesssim 0.8$. The value $\chi^2_{\rm red}\sim 1.1$ (p-value=0.38) in the plateau proves the appropriateness of the fitting function, Eq. \eqref{eq:H}. See the comments in the main text.}\label{fig:chi2}
\end{figure}

The reconstructions of $H(z)$ and $M(z)$ obtained with the baseline\_2D data set are presented in Fig. \ref{fig:H_M_2D_joined}. We set $z_{\rm max}=4$ and show again the 68\% of the total number of curves saved in the Monte Carlo Markov chain, only those with smallest $\chi^2$. In Fig. \ref{fig:chi2} one can see what is the decrease of the $\chi^2$ as a function of $z_{\rm max}$. The former remains stable for values of $z_{\rm max}\gtrsim 4$. Indeed, we find that in these cases the transition happens always at $0.5\lesssim z_t\lesssim 0.8$ and, hence, at a much higher redshift than in the case of the 3D BAO analysis studied in the preceding section, see again Fig. \ref{fig:post_zt}. It is easy to understand why. Transversal BAO data calibrated with $r_d^{P18}$ leads to smaller angular diameter distances than in $\Lambda$CDM (and larger values of $H(z)$) at $z\lesssim 0.7$. Thus, the shape of the Hubble function has to go below the standard one after these redshifts in order to compensate these effects and  respect the CMB preferred value of $D_A(z_*)$.

For sufficiently large values of $z_{\rm max}$ one can increase the compatibility with $M^{R22}$ within a larger redshift range, up to $z\gtrsim0.5$, and the deviation at higher redshifts cannot be considered to be statistically significant. The shapes of $H(z)$ and $M(z)$ leading to a potential solution to the Hubble tension are therefore quite different from those required by the 3D BAO data.  In this case we do not find significant departures of the deceleration parameter from the $\Lambda$CDM value at $z\ll1$. Again, our results remain stable under the inclusion of CCH data. 

\begin{figure}[t!]    
\includegraphics[scale=0.8]{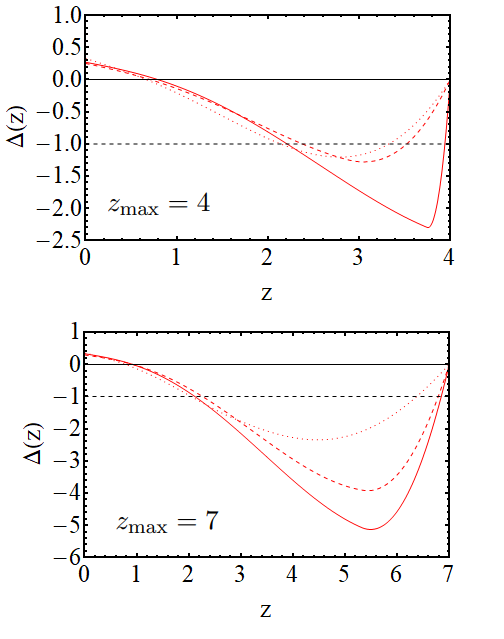}    \caption{Same as in Fig. \ref{fig:Delta3D}, but using in this case the baseline\_2D data set with $z_{\rm max}=4$ (upper plot) and $z_{\rm max}=7$ (lower plot). The transition happens in both cases at $z_t\lesssim 1$, with the dark energy density becoming negative ($\Delta<-1$) at $z\sim 2$. }\label{fig:Delta2D}
\end{figure}

If interpreted in terms of an effective self-conserved dark energy fluid, the change of $H(z)$ with respect to the $\Lambda$CDM requires in this case negative values of the dark energy density at $z\gtrsim 2$. The matter energy density at these redshifts is very large, so we need also a large value of $|\tilde{\rho}_{\rm de}|$ to have a sizable effect on the Hubble function, and $\tilde{\rho}_{\rm de}$ has to be negative in order the change to happen in the right direction. A couple of representative plots showing this characteristic behavior are presented in Fig. \ref{fig:Delta2D}. The crossing of the phantom divide is also allowed by the angular BAO data set.

Models with a negative DE density at these redshifts are available in the literature. Some examples are the sign-switching cosmological constant model of \cite{Akarsu:2019hmw,Akarsu:2021fol,Akarsu:2023mfb,Paraskevas:2023itu}, the self-conserved dark energy model of \cite{Gomez-Valent:2015pia} or models that consider dynamical dark energy on top of an anti-de Sitter vacuum with negative cosmological constant \cite{Visinelli:2019qqu,Calderon:2020hoc,Sen:2021wld,Adil:2023ara}\footnote{These models are encompassed inside the more general category of $\Lambda$XCDM models \cite{Grande:2006nn,Grande:2007wj,Grande:2008re}, which consider both a dynamical (running) $\Lambda$ term and a dynamical cosmon X with equation of state parameter different from -1, eventually interacting with $\Lambda$.}. In addition, it is interesting to note that several works have pointed to the possibility of the presence of unaccounted systematics in the standardization method of SNIa \cite{Kang:2019azh,Nicolas:2020lql,Lee:2021txi}\footnote{See e.g. Fig. 6 in \cite{Nicolas:2020lql}, in which the authors report an evolution of the mean SNIa stretch parameter as a function of redshift at $z<1$.}. This could potentially explain the smooth evolution of $M(z)$ hinted by our analysis when we consider the SH0ES prior and 2D BAO data is employed instead of 3D BAO to build the inverse cosmic ladder.


\subsection{WEC}\label{sec:WEC}

The results presented in Secs. \ref{sec:IDLanisotropic} and \ref{sec:IDLangular} tell us that all the low-$z$ solutions to the Hubble tension that involve new physics at $z\gtrsim 0.01$ require a violation of the weak energy condition, regardless of the BAO data set employed to build the inverse distance ladder. In both cases there must be a phantom-like evolution of the effective dark energy component, which manifests itself at very different moments of the cosmic expansion. Baseline\_3D requires this to happen at $z\lesssim 0.2$, whereas for Baseline\_2D the transition happens somewhere in the range $0.5\lesssim z\lesssim 0.8$. Moreover, in the latter case, the effective DE density takes negative values at $z\gtrsim 2$.

\begin{figure}[t!]
    \includegraphics[scale=0.39]{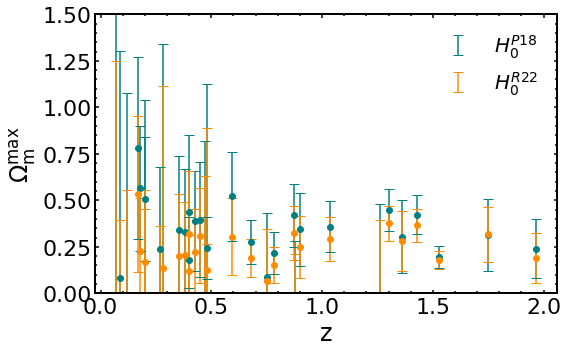}
    \caption{Upper bounds on the present value of $\Omega_m$ imposed by the WEC, see Eq. \eqref{eq:wec}. Only the (physical) range of positive values of $\Omega_m^{\rm max}$ is shown. We employ the method explained in Sec. \ref{sec:WECmethod}. Since the distributions for $\Omega_m^{\rm max}$ at some redshifts are highly non-Gaussian, we indicate their corresponding peaks (with dots) and the confidence intervals at 68$\%$ C.L. We show the results when $H_0$ is sampled using the {\it Planck}/$\Lambda$CDM value $H^{P18}_0$ (in blue) and the SH0ES measurement $H^{R22}_0$ (in orange).}
    \label{fig:wec}
\end{figure}

Now we apply the method of \cite{Sen:2007ep} (see Sec. \ref{sec:WECmethod}) to determine whether CCH, which are independent from the Baseline data sets, require the violation of the WEC given by Eq. \eqref{eq:wec}, assuming a self-conserved effective dark energy fluid. This condition must be obeyed if its energy density does not grow with the expansion, i.e. if DE is not phantom. We study how the result changes when we employ the SH0ES and {\it Planck} priors on $H_0$. The 33 constraints on $\Omega_m^{\rm max}$ obtained from the sampling of the CCH data listed in Table \ref{tab:CCH} and the priors on $H_0$ are shown in Fig. \ref{fig:wec}. In this calculation we have duly accounted for the correlations between the CCH data points. From the list of values of $\Omega_m^{\rm max}$ we extract a single representative upper bound on the matter parameter $\Omega_m$. Some of the error bars in Fig. \ref{fig:wec} are quite asymmetric, which means that the underlying multivariate distribution has some non-Gaussian features. Nevertheless, the deviation from Gaussianity in the case of the most precise values of $\Omega_m^{\rm max}$ is small, so we do not expect this to have a big effect on the final result. We will quantify its impact making use of the Edgeworth expansion, Eq. \eqref{eq:Edgeworth}, and will compare these results with those obtained assuming that the underlying distribution is a multivariate Gaussian with mean and covariance matrix given by Eq. \eqref{eq:Gapprox}.

Neglecting the non-Gaussian features we obtain, using the SH0ES and {\it Planck} priors on $H_0$, respectively, 

\begin{equation}\label{eq:Omax1}
\Omega_m^{\rm max}=0.250\pm 0.031\qquad [{\rm CCH}+H_0^{R22}]\,,
\end{equation}

\begin{equation}\label{eq:Omax2}
\Omega_m^{\rm max}=0.314\pm 0.036\qquad   [{\rm CCH}+H_0^{P18}]\,.
\end{equation}
We want to know whether current constraints on $\Omega_m$ fall below or above these upper bounds in order to determine whether they force the violation of the WEC, according to the CCH data set and the local values of $H_0$ considered in the computation of $\Omega_m^{\rm max}$. It is natural to check this for the value of $\Omega_m$ derived from CMB. If we consider standard physics before recombination, we can take {\it Planck}'s constraint $\omega_m^{P18}=0.1415\pm 0.0009$ and combine it with SH0ES and {\it Planck} priors on $H_0$, yielding

\begin{equation}\label{eq:O1}
\Omega_m=0.265\pm 0.008 \qquad [\omega_m^{P18}+H_0^{R22}]\,,
\end{equation}

\begin{equation}\label{eq:O2}
\Omega_m=0.315 \pm 0.007 \qquad [\omega_m^{P18}+H_0^{P18}] \,.
\end{equation}
These two results represent the constraints on $\Omega_m$ preferred by CMB, depending on whether we rely on a small or large value of the Hubble constant. To be fully consistent, we should compare these results with Eqs. \eqref{eq:Omax1} and \eqref{eq:Omax2}, respectively. Although the central values of $\Omega_m$ lie slightly above the upper bounds $\Omega_m^{\rm max}$, there is no important evidence for the violation of the WEC according to the CCH+$H_0^{R22}$ and CCH+$H_0^{P18}$ data sets if we assume standard prerecombination physics, since the values \eqref{eq:O1} and \eqref{eq:O2} fall below the upper bounds \eqref{eq:Omax1} and \eqref{eq:Omax2}, respectively, at $1\sigma$ C.L. This is at odds with the results reported in \cite{Gangopadhyay:2023nli}. 

\begin{figure}[t!]
    \includegraphics[scale=0.5]{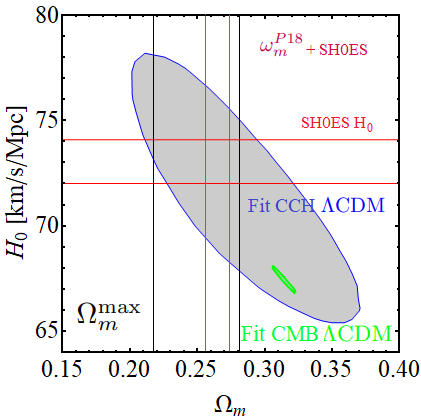}
    \caption{Contour plot at $68\%$ C.L. derived from the fit of the $\Lambda$CDM model to CCH and CMB data (in blue and green, respectively). We also include: (i) the constraint on $\Omega_m$ obtained by combining  $H_0^{R22}$ with the value of $\omega_m^{P18}$ inferred by {\it Planck} assuming standard prerecombination physics, Eq. \eqref{eq:O1} (in purple); (ii) the SH0ES measurement $H_0^{R22}$ (in red); and (iii) the upper bound on $\Omega_m^{\rm max}$, Eq. \eqref{eq:Omax1} (in black). All of them at 1$\sigma$ C.L.}
    \label{fig:contour}
\end{figure}

Our conclusions still hold true if we consider the non-Gaussian corrections of Eq. \eqref{eq:Edgeworth}, meaning that the impact of the non-Gaussian features in the distribution of $\Omega^{\rm max}_m$ is practically negligible, as expected. Indeed, we find {$\Omega^{\rm max}_m = 0.240 \pm 0.030$ and $\Omega^{\rm max}_m = 0.308 \pm 0.036$  at 68\% C.L. using the $H^{R22}_{0}$  and $H^{P18}_{0}$ priors, respectively. These results are fully compatible with those in Eqs. \eqref{eq:Omax1} and \eqref{eq:Omax2}.

In order to further illustrate all this we have fitted the flat $\Lambda$CDM model using CCH and provide the 68\% C.L. contours in Fig. \ref{fig:contour}\footnote{We obtain the following mean value and standard deviation: $H_0=(71.2\pm 4.2)$ km/s/Mpc and $\Omega_m=0.29\pm 0.06$.}. It is clear from that plot that, according to the data on cosmic chronometers, it is possible to explain a large value of $H_0\sim H_0^{R22}$ and a small value of $\Omega_m$ as the one in Eq. \eqref{eq:O1} within $\Lambda$CDM. In the standard model the WEC is automatically fulfilled due to the constancy of the DE density. Hence, we can explain the CCH data with the aforesaid values of $H_0$ and $\Omega_m$ without requiring phantom DE\footnote{A similar conclusion is reached in Ref. \cite{Sakr:2023hrl} by combining CCH with cluster counts and gas fraction in galaxy clusters, some astrophysical constraints on $\Omega_m$ with priors from Big Bang Nucleosynthesis measurements,
as well as from CMB correlations on $n_s$ and $A_s$.}. This is the same conclusion reached applying the method of \cite{Sen:2007ep}. The CCH+$H_0$ data themselves do not exclude the violation of the weak energy condition, but do not require its fulfilment either. At the moment, they cannot be used to strongly discriminate among possible solutions to the $H_0$ tension, and this is why the addition of CCH on top of Baseline\_2D and Baseline\_3D does not have a major impact on our results\footnote{See also \cite{Gomez-Valent:2021hda,Favale:2023lnp}, where it is shown that cosmic chronometers do not offer yet a very competitive calibration of the cosmic distance ladders. For instance, the reconstruction of $M(z)$ using CCH and SNIa is compatible with a constant, but does not exclude an evolution in redshift \cite{Favale:2023lnp}.}.

Fig. \eqref{fig:rho_bounds2} gives more support to our conclusions. We present there the constraints on the ratio $\rho_{\rm de}(z)/\rho_{\rm de}^0$ at 68\% C.L. obtained by using Eq. \eqref{eq:ratioDE} and sampling the CCH data together with the priors $\omega_m^{P18}$ and $H_0^{R22}$. Any of these measurements points to a clear violation of the WEC. We note, though, that the vast majority of the central points below $z=1$ fall in the range $0<\rho_{\rm de}(z)/\rho_{\rm de}^0<1$, which could be an indication of the preference of the data for a phantom behavior of the effective DE fluid. Moreover, the central point at $z=1.53$ is one sigma away from the positive region, and two sigma away from the border between the phantom and quintessence regions. This resonates very well with Fig. \ref{fig:wec} because the point at $z=1.5$ is the one that leads to the lowest upper bound on $\Omega_m^{\rm max}$, which is $\sim 2\sigma$ below the best-fit {\it Planck}/$\Lambda$CDM value, and hence in small tension with the standard model, as firstly noted in \cite{Sen:2007ep}. However, as already mentioned above, these hints of new physics are still mild.


\begin{figure}[t!]    
\includegraphics[scale=0.38]{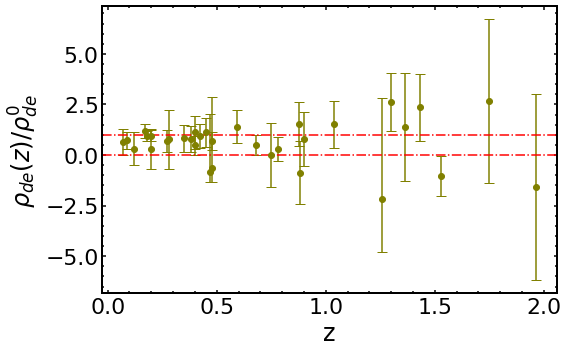}    \caption{Constraints on the ratio $\rho_{\rm de}(z)/\rho_{\rm de}^0$ obtained by using Eq. \eqref{eq:ratioDE} and sampling the CCH data together with the priors $\omega_m^{P18}$ and $H_0^{R22}$. We report the most probable values and 68\% confidence intervals. The dashed red lines at $0$ and $1$ denote the lower bounds for which the two conditions of Eq. \eqref{eq:WEC2} are satisfied, i.e. the positivity of the effective DE density and the non-phantom nature of the DE, respectively.}\label{fig:rho_bounds2}
\end{figure}

\section{Conclusions}\label{sec:conclusions}

The discovery of the cosmic acceleration in the late nineties meant a major breakthrough \cite{SupernovaSearchTeam:1998fmf,SupernovaCosmologyProject:1998vns}. It brought to the stage the need of adding a new component to the energy budget of the Universe, which must violate the strong energy condition. Its fundamental nature is largely unknown, but cosmological observations have let us infer some of its basic phenomenological properties, which are mimicked in the simplest scenario by a cosmological constant, see e.g. \cite{Padmanabhan:2002ji,Peebles:2002gy}. Despite the theoretical problems associated to it \cite{Weinberg:1988cp,Sola:2013gha}, it is a very important building block of the standard model of cosmology due to its good ability to fit the data. However, in the last decade, with the advent of precision cosmology, some mismatches between $\Lambda$CDM and 
 observations have irrupted into the scene \cite{Perivolaropoulos:2021jda}. The Hubble tension stands by far as the most significant one, since it already reaches the $\sim 5\sigma$ C.L. \cite{Planck:2018vyg,Riess:2021jrx}. Its solution could have serious theoretical implications, and this explains why understanding its origin has become one of the most pursued goals by the cosmological community \cite{Abdalla:2022yfr,Schoneberg:2021qvd,DiValentino:2021izs}.

In this paper we have devoted our efforts to study in detail the low-redshift phenomenology required to get rid of the $H_0$ tension, keeping standard physics before recombination. We have given special emphasis to the role played by the data on baryon acoustic oscillations, which is crucial in the construction of the cosmic inverse distance ladder. We have shown that anisotropic and angular BAO data (combined with CMB and SH0ES priors) lead to very different solutions, as expected \cite{Camarena:2019rmj}. The former require a phantom-like increase of the Hubble function and the absolute magnitude of supernovae of Type Ia at $0.01\lesssim z\lesssim 0.2$, whereas the latter (which in principle are less affected by model-dependent issues) need this increase to happen much earlier in the cosmic expansion, at $z_t\sim 0.5-0.8$, and more smoothly. $M(z)$ is in this case still compatible with a constant, but an evolution is not excluded\footnote{The evolution of $M(z)$ and the departure of $H(z)$ from its shape in the standard model could manifest themselves as an effective running of the $\Lambda$CDM parameters in low-$z$ analyses \cite{Millon:2019slk,Krishnan:2020vaf,Colgain:2022nlb}.}. In this scenario, if we consider that the new physics can be explained by an effective self-conserved dark energy component, the dark energy density has to be negative during, at least, some period of the cosmic history at $z\gtrsim 2$\footnote{It is useful to translate some of the most characteristic redshifts in this study into lookback time $t_{\rm lb}$: $z=(0.2,0.7,1,2)\to (2.4,6.3,7.7,10.2)$ Gyr. The relation $t_{\rm lb}(z)$ is a model-dependent quantity, of course. Here, for illustrative purposes, we have employed the flat $\Lambda$CDM with $H_0=70$ km/s/Mpc, $\Omega_m=0.3$ and three massless neutrinos.}. Hence, any low-$z$ solution to the Hubble tension with conserved effective dark energy demands a violation of the weak energy condition, and the possibility of a crossing of the phantom divide is not excluded. Coupled dark energy scenarios do not have to follow the conclusions that we have found assuming the conservation of DE, but have to give rise to forms of $H(z)$ and $M(z)$ compatible with our reconstructions. Another option to solve the Hubble tension is a local (ultra-late-time) change in $M$, in the second rung of the direct distance ladder, i.e. at $z\lesssim 0.01$, leaving intact the $\Lambda$CDM expansion history. Our results are in agreement with previous works that made use of anisotropic BAO data \cite{Alestas:2020zol} and also support the conclusions of \cite{Heisenberg:2022gqk,Heisenberg:2022lob}. Here, though, we make more definite statements than the latter, e.g. about the redshift ranges in which the phantom-like evolution should be active. They are extracted directly from the data. We also perform for the first time in the literature an analysis on similar lines using 2D BAO. 

In passing, we have also shown in several ways that current data on cosmic chronometers are not capable of determining whether the weak energy condition must be violated or not to solve the $H_0$ tension. They hint only very mildly to such a violation. Therefore, it is safe to assert that current CCH data do not help that much to constrain the form of the low-redshift solutions. This resonates well with \cite{Gomez-Valent:2021hda,Favale:2023lnp}.

The degree of naturalness of these solutions varies with the solutions themselves, but in this paper we wanted to focus on the phenomenology required to solve the Hubble tension and leave for future research the exploration of some of these routes in the context of concrete theoretical setups, considering also the evolution of perturbations and studying the symbiosis between the $H_0$ and growth tensions. The door for a late- and an ultra-late time solution to the Hubble tension is still open and, interestingly, the concrete form of the solutions depends crucially on the BAO data set that we consider. 

Future background and BAO data as those from {\it Euclid} \cite{EUCLID:2011zbd} are meant to be pivotal on the discussion and eventual solutions to the cosmic tensions. Particularly important will be the methods that the various collaborations employ to extract the information from the galaxy catalogs. We foresee the use of model-independent techniques to be relevant to obtain data sets as robust as possible, even if this comes at the expense of a decrease in precision \cite{Amendola:2019lvy,Amendola:2022vte}.


\vspace{0.25cm}
\noindent {\bf Acknowledgements}
\newline
\newline
\noindent AGV has been funded by the Istituto Nazionale di Fisica Nucleare (INFN) through the project of the InDark INFN Special Initiative: ``Dark Energy and Modified Gravity Models in the light of Low-Redshift Observations'' (n. 22425/2020). AGV, AF and MM acknowledge support by the INFN project “InDark”.  MM is also supported by the ASI/LiteBIRD grant n. 2020-9-HH.0 and by the  Fondazione  ICSC, Spoke 3 Astrophysics and Cosmos Observations, National Recovery and Resilience Plan (Piano Nazionale di Ripresa e Resilienza, PNRR) Project ID CN\_00000013 "Italian Research Center on High-Performance Computing, Big Data and Quantum Computing"  funded by MUR Missione 4 Componente 2 Investimento 1.4: Potenziamento strutture di ricerca e creazione di "campioni nazionali di R\&S (M4C2-19 )" - Next Generation EU (NGEU).
AAS acknowledges the funding from SERB, Govt of India under the research grant no: CRG/2020/004347.  AGV, AF and AAS acknowledge the participation in the COST Action CA21136 “Addressing observational tensions in cosmology with systematics and fundamental physics” (CosmoVerse). The authors thank the anonymous referee for their insightful comments and valuable suggestions.


\appendix

\section{Constraints in the $M$-$r_d$ plane from uncalibrated BAO and SNIa}\label{sec:appendixA}

We dedicate this appendix to explain how we obtained the degeneracy band in Fig. \ref{fig:degMrd}, which shows the big anticorrelation between the two calibrators of the direct and inverse distance ladders, $M$ and $r_d$, respectively. For this purpose, we make use of the so-called Index of Inconsistency (IOI) by Lin and Ishak \cite{Lin:2017ikq}. It is usually employed to quantify the level of inconsistency (or tension) between two data sets in the context of a concrete model when the posterior distributions of the parameters of the model are Gaussian in good approximation. However, it can also be employed to calibrate Gaussian data sets in a model-independent way \cite{Gomez-Valent:2021hda}. The IOI between two data sets takes the following form,

\begin{equation}\label{eq:IOI}
\rm{IOI}[i,j] = \frac{1}{2} \mu^{T} (C^{(i)} + C^{(j)})^{-1} \mu\,,
\end{equation}%
where $i$ and $j$ label the two data sets under consideration, whereas $\mu$ and $C$ denote, respectively, the difference between the corresponding data vectors and the covariance matrices. 

One can constrain the calibrators of two data sets by minimizing the IOI between them. Here, in particular, we want to study the correlation between $M$ and $r_d$ by minimizing the IOI between the SNIa and anisotropic BAO data sets. For each pair $(M,r_d)$, one can calibrate these data sets and extract measurements of angular diameter distances at several redshifts. In the case of SNIa, we do so by using Eq. \eqref{eq:M} and the Etherington relation \cite{Etherington:1933},

\begin{equation}\label{eq:Etherington}
D_A(z) = \frac{D_L(z)}{(1+z)^2}\,.
\end{equation}
In the case of BAO, we extract angular diameter distances from the ratios $D_A(z)/r_d$ and the dilation scales $D_V(z)/r_d$ collected in Table \ref{tab:BAO_data}\footnote{We do not employ here radial BAO data.}. For the latter, we use the expansion 

\begin{equation}\label{eq:DV_exp}
D_V \simeq \frac{3}{4}D_L\left(\frac{4}{3}z\right)\left(1+\frac{4}{3}z\right)^{-1}(1-0.0245z^{3}+0.0105z^{4})\,,
\end{equation}
which is quite accurate in the range $z<1$ for plausible accelerating cosmological models \cite{Sutherland_2012}. It allows us to compute the luminosity distance $D_L(4z/3)$ (and $D_A(4z/3)$, through Eq. \eqref{eq:Etherington}) from a measurement of $D_V(z)$. See also \cite{Lemos:2023qoy} for a recent application of this formula.

\begin{figure}[t!]    
\includegraphics[width=\columnwidth]{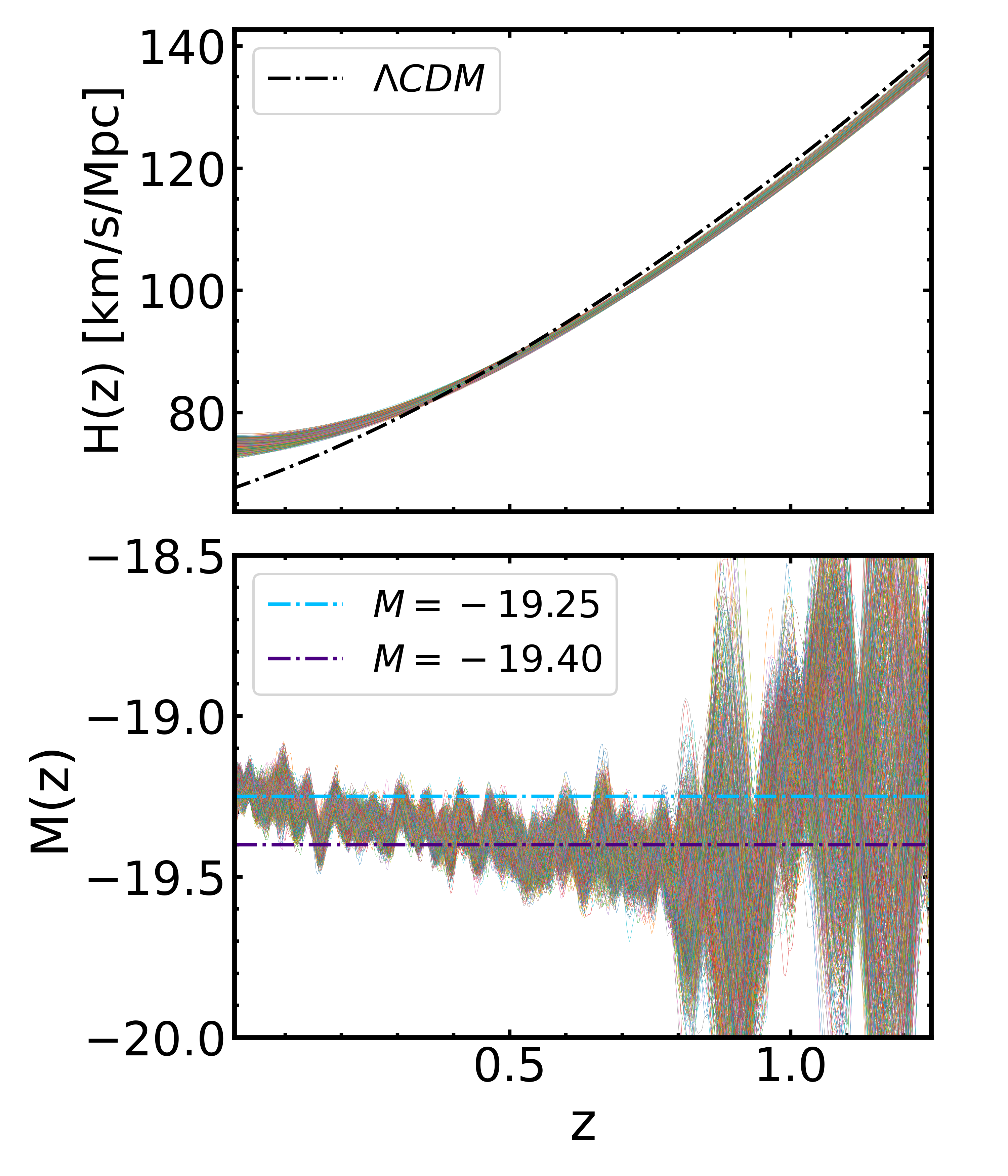}    \caption{Reconstructed shapes of $H(z)$ (upper plot) and $M(z)$ (lower plot) obtained with the Baseline\_2D data set and the parametrization of $\delta H _i(z)$ provided in Eq. \eqref{eq:deltaHalter}. We have set $z_{\rm max}=4$. See the comments in Appendix \ref{sec:appendixB}.}\label{fig:Hz_M_joined_newpar}
\end{figure}
We explore the space of the calibrators by employing the weights 

\begin{equation}\label{eq:weight}
    w=\exp(-{\rm IOI [BAO,SNIa]})\,,
\end{equation}
and a simple two-dimensional grid in the $M-r_d$ plane. Notice, though, that the data vectors entering Eq. \eqref{eq:IOI} need to correspond to the same redshifts. As this is not the case in the BAO and SNIa data sets, we choose to take the SNIa data points that fall right below and above the angular BAO data points in redshift and marginalize over the SNIa data that do not belong to this ensemble. Then, we sample the resulting distribution of SNIa apparent magnitudes $m(z)$ and compute the values at the BAO redshifts using a simple linear interpolation formula. This is licit because the difference in redshift of two consecutive SNIa points in the BAO redshift range is very small, so we can neglect the contribution of higher-order corrections. In this way, we end up with the central values of $m(z_{BAO})$ and the corresponding covariance matrix. This allows us to evaluate the weights using Eqs. \eqref{eq:IOI} and \eqref{eq:weight} and draw the distribution of $M$ and $r_d$, from which we get the contours at 68\% and 95\% C.L., see the grey band in Fig. \ref{fig:degMrd}.

\begin{figure}[t!]    
\includegraphics[scale=0.5]{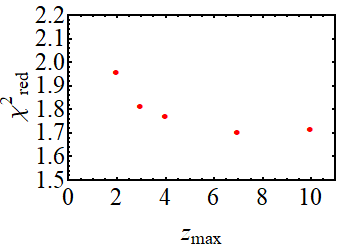}    \caption{Reduced $\chi^2$ as a function of $z_{\rm max}$ for the analyses with Baseline\_2D, using the parametrized form of $\delta H_i(z)$ given in Eq. \eqref{eq:deltaHalter}. As in Fig. \ref{fig:chi2}, $\chi^2_{\rm red}$ reaches a plateau. However, now $\chi^2_{\rm red}\sim 1.7$ instead of $\chi^2_{\rm red}\sim 1$, which means that the performance of Eq. \eqref{eq:deltaHalter} is much worse than Eq. \eqref{eq:deltaH}. See the comments in Appendix \ref{sec:appendixB}.}\label{fig:chi2Appendix}
\end{figure}


\section{Parameterization of $\delta H_i$ in terms on (1-a). Fitting formulae and results}\label{sec:appendixB}

In this appendix we test the robustness of the results of Secs. \ref{sec:IDLanisotropic} and \ref{sec:IDLangular} by comparing them with those obtained with an alternative fitting function. Instead of using Eq. \eqref{eq:deltaH}, here we employ 

\begin{eqnarray}\label{eq:deltaHalter}
\delta H_1(a) &= A+B(1-a)+C(1-a)^2\nonumber\\ \delta H_2(a) &= D+E(1-a)+F(1-a)^2\,,
\end{eqnarray}
with $\{A,B,C,D,E,F\}$ the new fitting parameters. The main reason in doing this is that expansion in terms of $(1-a)$ has better convergence compared to expansion in terms of $z$ and hence neglecting terms of the order $(1-a)^3$ and higher is better justified compared to neglecting terms of the order $z^3$ and higher. So the question is whether parametrization in terms of scale factor gives similar results as already obtained. The corresponding expressions in terms of the redshift are  obviously computed by doing $1-a=z/(1+z)$. 

Conditions \eqref{eq:cond1}-\eqref{eq:cond4} take now the following matrix form,

\begin{equation}
\begin{pmatrix}
\delta H_0 \\
\delta H_p \\
0\\
\end{pmatrix}= \begin{pmatrix}
1 & 0 & 0\\
1 & \frac{z_p}{1+z_p} & \frac{z_p^2}{(1+z_p)^2}\\
0 & 1 & \frac{2z_p}{1+z_p}
\end{pmatrix} 
\begin{pmatrix}
A \\
B \\
C\\
\end{pmatrix}
\end{equation}
and

\begin{equation}
\begin{pmatrix}
\delta H_{\rm max} \\
\delta H_p \\
0\\
\end{pmatrix}= \begin{pmatrix}
1 & \frac{z_{\rm max}}{1+z_{\rm max}} & \frac{z^2_{\rm max}}{(1+z_{\rm max})^2}\\
1 & \frac{z_p}{1+z_p} & \frac{z_p^2}{(1+z_p)^2}\\
0 & 1 & \frac{2z_p}{1+z_p}
\end{pmatrix} 
\begin{pmatrix}
D \\
E \\
F\\
\end{pmatrix}\,.
\end{equation}
The results obtained from the fitting analysis with the Baseline\_3D data set are essentially the same as those obtained with the fitting function of Eq. \eqref{eq:H}, see Sec. \ref{sec:IDLanisotropic}. The shapes of $H(z)$ and $M(z)$ have the same characteristic features shown in Fig. \ref{fig:H_M_3D_joined}. Thus, the conclusions of our main analysis with anisotropic BAO data hold true also for this new parametrization of the Hubble function.

The situation with the Baseline\_2D data is a bit different and deserves some detailed explanations. We present in Fig. \ref{fig:Hz_M_joined_newpar} the reconstructions of $H(z)$ and $M(z)$ obtained making use of Eq. \eqref{eq:deltaHalter}. We choose to fix $z_{\rm max}=4$ in this figure, but we have explicitly checked that the results remain stable for larger values of this parameter, as we found also in the analysis of Sec. \ref{sec:IDLangular}. This is reflected on the fact that the reduced $\chi^2$ reaches a plateau for $z_{\rm max}\gtrsim 4$, see Fig. \ref{fig:chi2Appendix}. However, now, in contrast to what we found with the parametrization \eqref{eq:deltaH}, the transition happens at smaller redshifts, at $z_t\sim 0.4$ (instead of $z_t\sim 0.5-0.8$). This is true for both, the Hubble parameter and the absolute magnitude of SNIa. Nevertheless, it is very important to notice that the goodness of fit found with the new parametrization is much worse than the one offered by the original parametrization. Notice that $\chi^2_{\rm red}\sim 1.7$ in the plateau of Fig. \ref{fig:chi2Appendix}. This corresponds to a quite small p-value=0.03. Conversely, we obtained $\chi^2_{\rm red}\sim 1.1$ (p-value=0.38) with parametrization \eqref{eq:deltaH}  although the results are similar in both parametrizations. This justifies the use of the latter (in terms of $z$) in the main body of the paper.


\section{Other tests on the robustness of our parametrization of $H(z)$}\label{sec:appendixC}
We have performed some additional tests to assess the robustness of the results obtained with the fitting function \eqref{eq:H}. We perform all of them using the Baseline\_3D data set.

To begin with, we consider the following variant,

\begin{equation}
H(z) = \left\{
        \begin{array}{ll}
             H_\Lambda(z)+ \delta H_1(z)& {\rm if}\quad 0<z \leq z_{p} \\
             H_\Lambda(z)+ \delta H_2(z)& {\rm if}\quad z_p<z \leq z_{\rm max} \\
            H_\Lambda(z)& {\rm if} \quad z \geq z_{\rm max}
        \end{array}
    \right.
\end{equation}
in which we substitute $\bar{H}(z)$ by $H_\Lambda(z)$ at $z<z_{\rm max}$. The results remain completely stable, with no apparent difference with respect to those obtained in the main analysis. The values of $\chi^2_{\rm min}$ and the reconstructed shapes of $H(z)$ and $M(z)$ are basically indistinguishable from those reported in Sec. \ref{sec:IDLanisotropic}.

\begin{figure*}[t!]    
\includegraphics[scale=0.55]{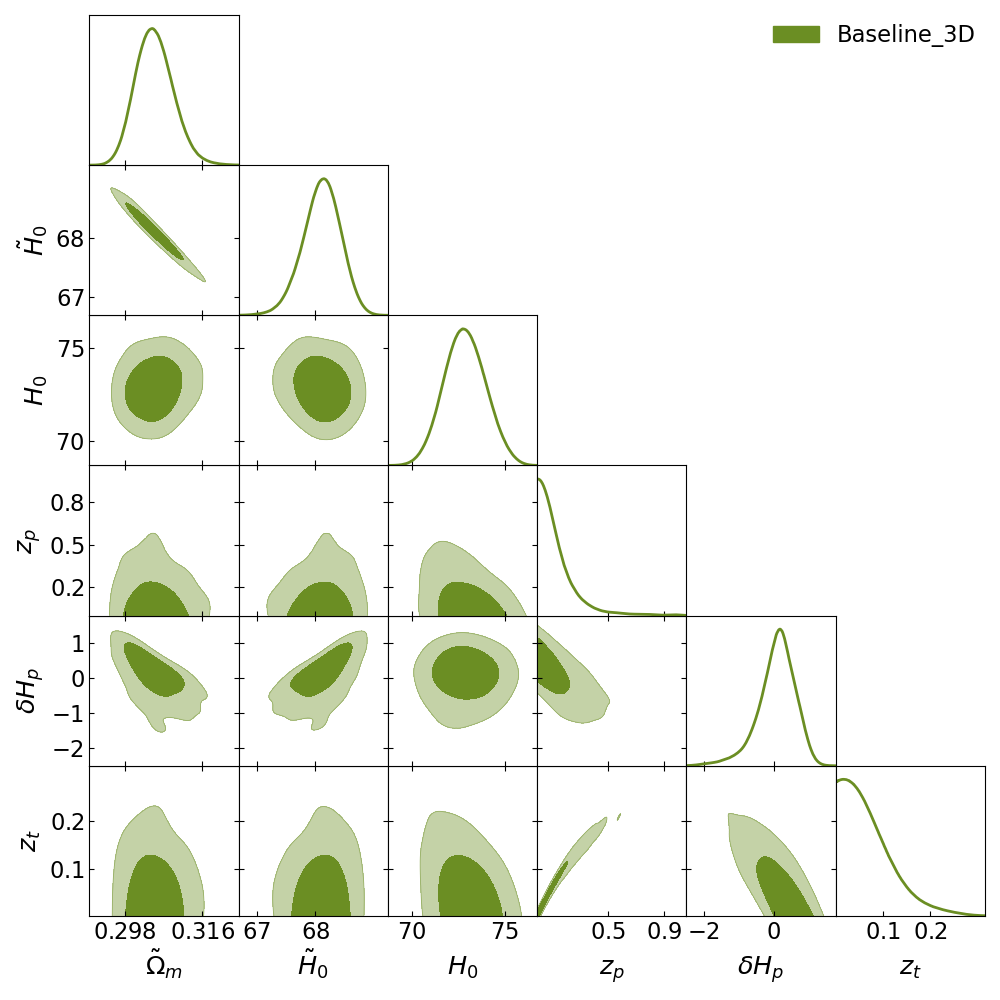}    \caption{Triangle plot with the constraints obtained for the various fitting parameters entering the Baseline\_3D analysis and the derived parameter $z_t$, setting $z_{\rm max}=1$. The triad $\{\tilde{H}_0,H_0,\delta H_p\}$ is given in units of km/s/Mpc. Contours contain 68\% and 95\% of the probability.}\label{fig:triangle_3D}
\end{figure*}

We then study what happens if instead of forcing the functions $\delta H_1$ and $\delta H_2$ to have a minimum at $z_p$, we still force the continuity of $dH_p/dz$ at $z_p$, but allow its value to vary freely in the Monte Carlo. The value of $\chi^2_{\rm min}$ decreases $\sim 1$ unit. However, in this case we have one more degree of freedom, so from a Bayesian perspective this more general version of the fitting function is not particularly favored despite the slight improvement in the description of the data. Indeed, we have checked that the posterior distribution of $dH/dz|_{z=z_p}$ encompasses the 0 value at $<1\sigma$ C.L. Something similar happens if we force the continuity of the second derivative of the Hubble function, instead of the continuity of the first derivative.

Finally, in our last test we have considered

\begin{equation}
H(z) = \left\{
        \begin{array}{ll}
             \bar{H}(z)+ \delta H(z)& {\rm if}\quad 0<z \leq z_{\rm max} \\
            H_\Lambda(z)& {\rm if} \quad z \geq z_{\rm max}
        \end{array}
    \right.
\end{equation}
with

\begin{equation}
\delta H(z) = c_0+c_1z+c_2z^2+c_3z^3
\end{equation}
satisfying the boundary condition $\delta H(z_{\rm max})=H_\Lambda(z_{\rm max})-\bar{H}(z_{\rm max})$. Its performance is much worse than the one employed in the main analysis. We find an increase $\Delta\chi^2_{\rm min}\approx +6$. This fitting function cannot fully accommodate simultaneously the SH0ES and {\it Planck} distance priors and, hence, is not suited for the aim of this work, namely the study of the low-redshift solutions to the Hubble tension.


\begin{table}[htbp!] 
\centering
\renewcommand{\arraystretch}{2}
\begin{tabular}{|c ||c | c |}
\hline
{\small Parameter} & {\small Baseline\_3D}  & {\small Baseline\_2D} 
\\\hline
$\tilde{\Omega}_m$ & $ 0.305^{+0.004}_{-0.005}$ & $ 0.313^{+0.008}_{-0.007}$  \\\hline
$\tilde{H}_0$ [km/s/Mpc] & $ 68.11^{+0.34}_{-0.29}$ & $ 67.51^{+0.55}_{-0.49}$ \\\hline
$H_0$ [km/s/Mpc] & $ 72.90 \pm 1.10 $ & $ 73.97^{+0.76}_{-0.69} $ \\\hline
$\delta H_p$ [km/s/Mpc] & $ 0.08^{+0.58}_{-0.40} $ & $ -8.90^{+2.20}_{-2.60}$ \\\hline
$z_p$ & $ <0.41 $ & $ 3.07^{+0.87}_{-0.44}$ \\\hline\hline
$z_t$ & $ <0.17 $ & $ 0.66^{+0.07}_{-0.04} $ \\\hline\hline
$\chi^2_{\rm min}$ & $16.14$ & $16.69$
\\\hline
p-value& $0.21$ & $0.38$\\
\hline
\end{tabular}
\caption{Mean values and 68\% C.L. errors for the main and derived parameters obtained  from the analyses with Baseline\_3D and Baseline\_2D. The upper bounds, instead, are given at 95\% C.L. In the last two rows we report the best-fit $\chi^2$ and the corresponding p-values. The number of degrees of freedom is 12 and 14 in the Baseline\_3D and Baseline\_2D analyses, respectively.}
\label{tab:table_fits}
\end{table}

\section{Triangle plots}\label{sec:appendixD}

We devote this appendix to show the one-dimensional posterior distributions and two-dimensional contour plots of the fitting parameters $\{\tilde{\Omega}_m,\tilde{H}_0,H_0,z_p,\delta H_p\}$ and the transition redshift $z_t$ obtained from the analyses of the Baseline\_3D and Baseline\_2D data sets in Figs. \ref{fig:triangle_3D} and \ref{fig:triangle_2D}, respectively. For completeness, in Table \ref{tab:table_fits} we provide all the mean values and 68\% C.L. uncertainties. We have obtained these results using \texttt{GetDist} \cite{Lewis:2019xzd}. As discussed in the main text, the constraints on $\tilde{\Omega}_m$ and $\tilde{H}_0$ are driven by the {\it Planck}/$\Lambda$CDM prior, while the constraint on $H_0$ is essentially fixed by the SH0ES prior. Conversely, the results for $\delta H_p$ and $z_p$ (and $z_t$) depend on the data set under study, and we find in both cases, as expected, a strong positive correlation between $z_t$ and $z_p$ and an anti-correlation between $z_p$ and $\delta H_p$.

\begin{figure*}[t!]    
\includegraphics[scale=0.55]{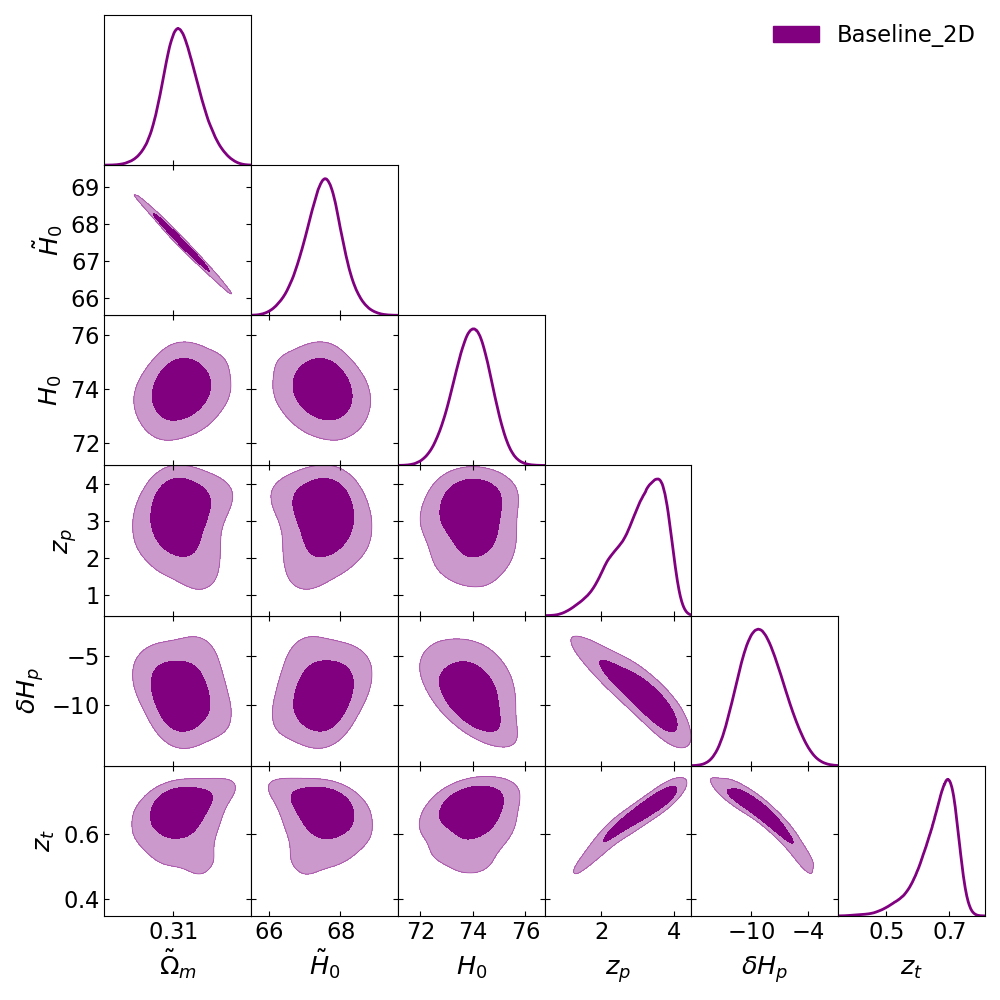}    \caption{Same as in Fig. \ref{fig:triangle_3D}, but for the Baseline\_2D analysis, with $z_{\rm max}=4$.}\label{fig:triangle_2D}
\end{figure*}


\clearpage

\bibliographystyle{apsrev4-1}
\bibliography{lowzSolH0}

\end{document}